\documentclass[manuscript]{aastex}

\slugcomment{Not to appear in Nonlearned J., 45.}

\shorttitle{A New Method To Classify Flares Of UV Ceti Type Stars}
\shortauthors{Dal and Evren}

\begin{document}


\title{A New Method To Classify Flares Of UV Ceti Type Stars: \\
Differences Between Slow And Fast Flares}


\author{H. A. Dal and S. Evren}
\affil{Department Of Astronomy and Space Sciences, University of Ege, \\
Bornova, 35100 ~\.{I}zmir, Turkey}

\email{ali.dal@ege.edu.tr}



\begin{abstract}

In this study, a new method is presented to classify flares derived from the photoelectric photometry of UV Ceti type stars. Using \textit{Independent Samples t-Test}, the method is based on statistical analysis. The data used in the analyses were obtained from four flare stars observed between the years 2004 and 2007. Total number of flares obtained in the observations of \object{AD Leo}, \object{EV Lac}, \object{EQ Peg} and V1054 Oph is 321 in the standard Johnson U band. As a result, flare can be separated into two types as \textit{slow} and \textit{fast} depending on the ratio of \textit{flare decay time} to \textit{flare rise time}. The ratio is below the value 3.5 for all slow flares, while it is above 3.5 for all fast flares. Also, according to the \textit{Independent Samples t-Test}, there are about 157 seconds difference between equivalent durations of slow and fast flares. In addition, there are significant differences between amplitudes and rise times of slow and fast flares.

\end{abstract}




%% Otherwise, LaTeX will issue an error. 

\section{Introduction}

Flares and flare processes are very hard worked subtitles of astrophysics. Lots of studies on flares have been carried out since the first flare was detected on the Sun by R.C. Carrington and R. Hodgson in 1 September 1959. Flare processes have not been perfectly understood yet \citep{Ben10}. However, the researches indicate that the incidence of red dwarfs in our galaxy is 65$\%$. Seventy-five per cent of them show flare activity, these stars are known as UV Ceti type stars \citep{Rod86}. In this respect, it will be easier to understand the evolution of red dwarfs if the flare processes are well known. This is because flare activity dramatically affects the evolutions of the red dwarfs. In this respect, an attempt to classify flares by considering shapes of flare light variations observed in the UV Ceti type stars, flares have been tried to classify. It is believed that classification of flares makes the flares and flare processes more intelligible.

Flares of the UV Ceti type stars were first detected in 1939 \citep{Van40}. Discovering the flare stars with high flare frequency such as UV Cet, YZ CMi, \object{EV Lac}, \object{AD Leo} and \object{EQ Peg}, the detected flare numbers and the variety of flare light variations increased. The light variation of each flare is almost different from each other. In the first place, it is seen that there are lots of shapes for flare light variations \citep{Mof74, Ger05}. On the other hand, when large numbers of flares are examined, it is seen that there are only two main shapes for flare light variations. One of them is called the fast flare. Fast flares have higher energy and the shapes of their light variations are similar to the shapes of solar hard X-ray flares. The second type flares are called slow flares. Unlike fast flares, slow flares exhibit lower energy. The rise times of slow flares are almost equal to their decay times.

The terms of fast and slow flares were used for the first time in 1960's in astrophysics. If the rise time of a flare is below 30 minutes, \citet{Har69} called that flare \textit{a fast flare}. If its rise time is above 30 minutes, they called it \textit{a slow flare}. Like \citet{Har69}, considering the shapes of flare light variations, \citet{Osa68} described two types of flares. However, \citet{Osk69} separated flares into four classes. Like \citet{Osk69}, considering only light variation shapes of the flares, Moffett (1974) directly classified flares such as classical, complex, slow, and flare event. On the other hand, Kunkel had asserted another idea in his PhD thesis in 1967 \citep{Ger05}. According to Kunkel, the observed flare light variations must be some combinations of some slow and fast flares. According to this idea, there are two main flare types. The complex flares mentioned by \citet{Mof74} are actually a combination of some fast and slow flares. And also, both slow flares and flare events mentioned by \citet{Mof74} can be classified as the same type flares.

\citet{Gur88} described two flare types to model the flare light curves. \citet{Gur88} indicated that thermal processes are dominated in the processes of slow flares. And these flares are ninety-five per cent of all flares observed in UV Ceti type stars. The non-thermal processes are dominated in the processes of fast flares, which are all the other flares. According to \citet{Gur88}, there is a large energy difference between these two types of flares.

In this study, we introduce a new statistical method for classifying flares. Using statistical \textit{Independent Samples t-Test} (hereafter \textit{t-Test}) analysis, this method is based on the distribution of flare equivalent durations versus flare rise times. Considering the studies in \citet{Osa68}, Kunkel's PhD thesis \citep{Ger05} and \citet{Gur88}, we assume that there are two flare types such as fast and slow flares types. We classify flares into two types and demonstrate the similarities and differences between these two types of flares. In respect of these analyses, we discuss the results obtained from analyses of 321 flares detected in U band observations of flare stars \object{AD Leo}, \object{EV Lac}, \object{EQ Peg} and \object{V1054 Oph} between 2004 and 2007. The programme stars were selected for this study due to their high flare frequencies \citep{Mof74}. The flare data obtained in this study are useful for such analysis. This is because the data were obtained with systematical observations, using the same method.

The flare activity of \object{AD Leo} was discovered by \citet{Gor49} for the first time. The star is a red dwarf and a member of The Castor Moving Group, whose age is about 200 million years \citep{Mon01}. \citet{Cre06} found that the flare frequency of \object{AD Leo} was 0.71 $h^{-1}$. The variation of the flare frequency was investigated by \citet{Ish91}. They mentioned that there is no variation in the flare frequency of \object{AD Leo}. The other star in this study is \object{EV Lac}, which is one of the well known UV Ceti stars. According to the spatial velocities, \object{EV Lac} seems to be a member of 300 million years old Ursa Major Group \citep{Mon01}. It has been known since 1950 that \object{EV Lac} shows flares \citep{Lip52, Van53}. The largest observed flare amplitude is 6.4 mag in U band observations of \object{EV Lac}. \citet{And82} detected 50 flares in U and B band observations of \object{EV Lac}. The author indicated that about 42 flares of 50 flares were found in some groups, which were occurred in every 5-6 days. The seasonal flare frequencies were computed from 1972 to 1981 and these frequencies were compared with the seasonal averages of B band magnitudes. According to this comparison, it was found that the activity cycle is about 5 years for \object{EV Lac} \citep{Mav86}. On the other hand, \citet{Ish91} indicated that there was no flare frequency variation from 1971 to 1988. In another study, \citet{Let97} showed that flare frequencies of \object{EV Lac} increased from 1968 to 1977. \object{EQ Peg} is another active flare star, whose flare activity was discovered by \citet{Roq54}. \object{EQ Peg} is classified as a metal-rich star and it is a member of the young disk population in the galaxy \citep{Vee74, Fle95}. \object{EQ Peg} is a visual binary \citep{Wil54}. Both of the components are a flare star \citep{Pet83}. Angular distance between components is given as a value between 3$\arcsec$.5 and 5$\arcsec$.2 \citep{Hai87, Rob04}. One of the components is 10.4 mag and the other is 12.6 mag in V band \citep{Kuk69}. Observations show that flares of \object{EQ Peg} generally come from the fainter component \citep{Fos95}. \citet{Rod78} proved that 65$\%$ of the flares come from faint component and about 35$\%$ from the brighter component. The fourth star in this study is V1054 Oph, whose flare activity was discovered by \citet{Egg65}.  V1054 Oph (= Wolf 630ABab, Gliese 644ABab) is a member of Wolf star group \citep{Joy47, Joy74}. Wolf 630ABab, Wolf 629AB (= Gliese 643AB) and VB8 (= Gliese 644C), are the members of the main triplet system, whose scheme is demonstrated in Fig.1 given in the paper of \citet{Pet84}. Wolf 630 and Wolf 629 are visual binary and they are separated 72$\arcsec$ from each other. Wolf 630AB is a close visual binary in itself. Wolf 629AB is a spectroscopic binary. B component of Wolf 629AB seems to be a spectroscopic binary. VB8 is 220$\arcsec$ far away from the other components. There is an angular distance about 0$\arcsec$.218 between A and B components of Wolf 630 \citep{Joy47, Joy74}. The masses were derived for each components of Wolf 630ABab by \citet{Maz01}. The author showed that the masses are 0.41 $M_{\odot}$ for Wolf 629A, 0.336 $M_{\odot}$ for Wolf 630Ba and 0.304 $M_{\odot}$ for Wolf 630Bb. In addition, \citet{Maz01} demonstrated that the age of the system is about 5 Gyr.

\section{Observations and Analyses}

\subsection{Observations}

The observations were acquired with a High-Speed Three Channel Photometer attached to a 48 cm Cassegrain type telescope at Ege University Observatory. Using a tracking star set in a second channel of photometer, the observations were continued in standard Johnson U band with the exposure time between 2 and 10 seconds. The basic parameters of all program stars and their comparisons are given in Table 1. The parameters given in the table are star name, magnitude in V band, B-V colour index, spectral type, distance (pc) and bolometric luminosity ($LogL_{bol}$, ergs $s^{-1}$). The magnitudes and colour indexes were obtained in this study. Considering B-V colour indexes, the spectral types were taken from \citet{Tok00}. The distances and bolometric luminosities were taken from \citet{Fos95} and \citet{Ger99}. 

Although the program and comparison stars are so close on the sky, differential extinction corrections were applied. The extinction coefficients were obtained from the observations of the comparison stars on each night. Moreover, the comparison stars were observed with the standard stars in their vicinity and the reduced differential magnitudes, in the sense variable minus comparison, were transformed to the standard system using procedures outlined in \citet{Har62}. The standard stars are listed in the catalogues of \citet{Lan83, Lan92}. Heliocentric corrections were applied to the times of the observations. The standard deviation of observation points acquired in standard Johnson U band is about 0$^{m}$.015 on each night. Observational reports of all program stars are given in the Table 2. It is seen that there is no variation of differential magnitudes in the sense comparison minus check stars.

\citet{Ger72} developed a method for calculating flare energies. Flare equivalent durations and energies were calculated with using Equation (1) and (2) of this method. 

\begin{center}
\begin{equation}
P~=~\int[(I_{flare}~-~I_{0})~/~I_{0}]~dt
\end{equation}
\end{center}
In Equation (1), $I_{0}$ is the intensity of the star in quiescent level. $I_{flare}$ is the intensity during flare.

\begin{center}
\begin{equation}
E~=~P~\times~L
\end{equation}
\end{center}
where $E$ is the flare energy; $P$ is the flare equivalent duration; $L$ is the luminosity of the stars in quiescent level in the Johnson U band.

HJD of flare maximum times, flare rise and decay times, amplitudes of flares and flare equivalent durations were calculated for each flare. Brightness of a star without a flare was taken as a quiescent level of brightness of this star on each night. Considering this level, all flare parameters were calculated for each night. It was seen that some flares have a few peaks. In this case, flare maximum time and amplitudes were calculated from the first highest peak. Instead of flare energies, flare equivalent durations were used for all statistical analyses. This is because of the luminosity term in Equation (2). The luminosities of stars from different spectral types have great differences. Although the equivalent durations of two flares obtained from two stars in different spectral types are the same, calculated energies of these flares are different due to different luminosities of these spectral types. Therefore, we could not use these flare energies in the same analyses. On the other hand, flare equivalent duration depends just on flare power. Another reason of using equivalent duration is that the given distances of the same star in different studies are quite different. Therefore, the calculated luminosities become different because of these different distances.

All the calculated parameters of flares are given in Table 3. The given parameters in columns are star name, observation date, HJD of flare maximum, flare total time (s), decay time (s), equivalent duration (s), energy (ergs), flare amplitude (mag) and flare types. In the last column, it is given whether the flare was used in the analyses, or not.

When the observed flares are examined, it is seen that almost each flare has a distinctive light variation shape (Figures 1, 2, 3, and 4). In these figures, horizontal dashed lines represent the level of quiescent brightness. The flare seen in Figure 1 is a fast flare. This flare type occurs frequently in UV Ceti type stars. On the other hand, the first flare seen in Figure 2 is a compact flare. This flare type among the others is the hardest flare type to classify. This flare type must be a combination of two flares. The observed flares, whose light variations are similar to the first flare in Figure 2, were not used in the analyses. The second flare seen in Figure 2 is a fast flare. The flare seen in Figures 3 is a powerful flare, but its light variation was not completed because we did not carry on observation until the flare completely decreased to quiescent phase due to the Sun rising. If the light variation of a flare was not completed like this one, the flare was not used in the analyses. The flare seen in Figure 4 seems to be very different from previous flares. \citet{Mof74} called flares like this one as the flare events. They are called slow flares in some other studies. In this study, we called them as the slow flares.

\subsection{Analyses}

The impulsive phase of a flare is the time interval where sudden-high energy occurs. On the other hand, the mean phase is other part of the flare, where the energy is emitted to all space \citep{Gur88, Ben10}. Moreover, rate of brightness increase for fast flares is clearly higher than that of slow flares \citep{Gur88, Ger05}. Moreover, \citet{Gur88} stated that, the energies of fast flares are always higher than slow flare energies.

According to this approach, we examined all flare data and we saw that the equivalent durations of some flares are different, while their rise times are the same. For example, the rise time of 22 flares is 15 second. The equivalent durations of eight flares among them are very high, while the equivalent durations of other 14 flares are dramatically low. It was seen in 30 different rise times. It means that there are least two flares in 30 different rise times and their equivalent durations are different from each other. 140 flares were chosen in total. These 140 flares used in the analyses are specified in the last column of Table 3. Considering also their light variation shapes, we separated flares into two groups as flares with high energy and low energy. It was found that 61 flares have high energy, while 79 flares have low energy. Considering light variations, it was seen that 61 flares are fast flare, and the others are slow flare. For each one of 30 rise times, the averages of the equivalent durations were computed separately for the flares with high energy and low energy.

The most suitable statistical test for these data is the \textit{t-Test} to determine the difference between the equivalent durations of two groups. This is because the \textit{t-Test} examines whether there is any statistical difference between independent variable of two groups, or not \citep{Wal03, Daw04}. In this study, the flare rise times were taken as a dependent variable, while flare equivalent durations were taken as an independent variable. In the analyses, the SPSS V17.0 software was used \citep{Gre99}. The average of equivalent durations in the logarithmic scale for 79 slow flares was calculated and found to be 1.348 $\pm$ 0.092. And it was found to be 2.255 $\pm$ 0.126 for 61 fast flares. This shows that there is a difference about 0.907 between average equivalent durations in the logarithmic scale. The \textit{p-probability value} (hereafter \textit{p-value}) was computed to test the results of the \textit{t-Test} and it was found as \textit{p-value} $<$ 0.0001. It means that the result is statistically acceptable. All the results of the analyses are given in Table 4.

In the next step, we compared their distributions, the distribution of the equivalent durations versus flare rise times. The best fits for distributions were searched. Using GrahpPad Prism V5.02 software \citep{Mot07}, regression calculations showed that the best fits of distributions seen in Figure 5 were linear functions given by Equations (3) and (4).

\begin{center}
\begin{equation}
Log(P_{u})~=~1.109~\times~Log(T_{r})~-~0.581
\end{equation}
\end{center}

\begin{center}
\begin{equation}
Log(P_{u})~=~1.227~\times~Log(T_{r})~+~0.122
\end{equation}
\end{center}

It was tested whether these linear functions belong to two independent distributions, or not. In this point, the slopes of linear functions were principally examined. As it can be seen in Table 4, the slope of the linear function is 1.109 $\pm$ 0.127 for slow flares, while it is 1.227 $\pm$ 0.243 for fast flares. This shows that the increasing of equivalent durations versus flare rise times for both fast and slow flares are parallel. When the probability, \textit{p-value}, was calculated to say whether it is statistically significant, it was found that \textit{p-value} = 0.670. This value indicates that there is no significant difference between the slopes of fits.

Finally, the \textit{y-intercept} values were calculated and compared for two linear fits. While this value is -0.581 for the slow flares, it is 0.122 for the fast flares in the logarithmic scale. It means that there is a difference about 0.703 between these values in the logarithmic scale. When the probability value was calculated for \textit{y-intercept} values to say whether there is a statistically significant difference, it was found that \textit{p-value} $<$ 0.0001. This result indicates that the difference between two \textit{y-intercept} values is clearly important.

Some other differences like ones demonstrated by \textit{t-Test} are directly seen in the graphics. For example, the lengths of flare rise times for both type flares can be compared in Figure 6. While the lengths of rise times for slow flares can reach to 1400 seconds, they are not longer than 400 seconds for fast flares.

The comparison of another parameter is given in Figure 7. The flare amplitudes are seen in this figure. As it can be seen, while the amplitudes of fast flares can reach to 4.0 mag, the amplitudes of slow flares can exceed 1.0 mag.

61 fast and 79 slow flares were chosen among 321 flares observed in this study. The ratios of flare decay time to flare rise time were computed for both 61 fast and 79 slow flares. As a result, it is seen that the ratio is below the value of 3.5 for each one of 79 slow flares. On the other hand, the ratio is above the value of 3.5 for each fast flares. The value of 3.5 is considered as a limit for these two type flares. Considering the ratio of 3.5, other 181 flares of 321 flares were separated as slow and fast flares. When the results obtained from analyses of 140 flares were rechecked for 321 flares, it was seen that the results are the same with the previous ones.

\section{Results and Discussion}

We observed 321 flares in U band observations of \object{AD Leo}, \object{EV Lac}, \object{EQ Peg} and V1054 Oph. Examining 321 flares, 61 fast and 79 slow flares were identified for analyses. The \textit{t-Test} was used as an analysis method. Flare rise times were accepted as dependent variables, while flare equivalent durations were taken as independent variables. The results obtained from the \textit{t-Test} analyses of the data show that there are distinctive differences between two flare types. These differences are important properties because the models of white light flares observed in photoelectric photometry must support these properties to explain both flare types.

The distributions of the equivalent durations were represented by linear fits given by Equations (3) and (4) for these flare types. The slope of linear fit is 1.109 for slow flares, which are low energy flares. And, it is 1.227 for fast flares, which are high energy flares. The values are almost close to each other. It seems that the equivalent durations versus rise times increase in similar ways.

In the case of UV Ceti stars, when flare models are considered, it is seen that there are two main energy sources for flares \citep{Gur88, Ben10}. These depend on the thermal and non-thermal processes \citep{Gur88}. Flares with small amplitude are generally the flares with low energy. The thermal processes are commonly dominant for these type flares. On the other hand, the flares, which have sudden rapid increases, are more energetic events. Non-thermal processes are dominant for this type. And thus, there is an energy difference between these two types of flares \citep{Gur88}. When the averages of equivalent durations for two type flares were computed in logarithmic scale, it was found that the average of equivalent durations is 1.348 for slow flares and it is 2.255 for fast flares. The difference of 0.907 between these values in logarithmic scale is equal to 157.603 second difference between the equivalent durations. As it can be seen from Equation (2), this difference between average equivalent durations affects the energies in the same way. Therefore, there is 157.603 times difference between energies of these two type flare. This difference must be the difference mentioned by \citet{Gur88}. 

The slopes of linear fits are almost close. On the other hand, if the \textit{y-intercept} values of the linear fits are compared, it is seen that there is 0.703 times difference in logarithmic scale, while there is 0.907 times difference between general averages. Considering also Figure 5, it is seen that equivalent durations of fast flares can increase more than slow flare equivalent durations towards the long rise times. Some other effects should be involved in the fast flare process for long rise times. These effects can make fast flares more powerful than they are.

When the lengths of rise times for both flare types are compared, it is seen that there is a difference between them. The lengths of rise times can reach to 1400 seconds for slow flares, but are not longer than 400 seconds for fast flares. In addition, when the flare amplitudes are examined for both type flares, an adverse difference is seen according to rise times. While the amplitudes of slow flares reach to 1.0 mag at most, the amplitudes of fast flares can exceed 4.0 mag.

Finally, when the ratios of flare decay times to flare rise times are computed for two flares types, the ratios never exceed the value of 3.5 for all slow flares. On the other hand, the ratios are always above the value of 3.5 for fast flares. It means that if decay time of a flare is 3.5 times longer than its rise time at least, the flare is a fast flare. If not, the flare is a slow flare. Therefore, the type of an observed flare can be determined by considering this value of the ratio. In the studies like \citet{Osa68}, \citet{Osk69}, \citet{Har69} and \citet{Mof74}, considering directly the shapes of flare light variations, the flares have been classified into two types as fast and slow flares. For instance, according to \citet{Har69}, if the rise time of a flare is above 30 minutes, the flare is slow flare. If not, it is a fast flare. However it is shown in this study that there are some fast flares, whose rise times are longer than the rise times of some slow flares. This is clearly seen from Table 3. This case indicates that a classification by considering only the rise time may not be right. Nevertheless, \citet{Mof74} separated flares into more than two groups such as classic, complex, spike and flare events. On the other hand, according to our results of \textit{t-Test} analyses, neither only one parameter nor the shape of the light variation was enough to classify a flare. The flare equivalent durations and also one more parameter should be taken into consideration in order to make such a classification.

The values 3.5, the ratio of flare decay times to flare rise times, can give an idea about the rate of energy emitting in a flare process. The rise times of flares are some limits for each type. Maximum flare rise time is about 400 seconds for fast flares, while it can reach the values over 1400 seconds for slow flares. However, the decay times can take any duration without any limited values. Consequently, the ratio of flare decay times to flare rise times depends on rise time more than decay times. In the case of rise time, the difference between two type flares must be caused by whether the flare processes are thermal or non-thermal. We computed the duration as a rise time from the phase in which the brightness increases. Increasing of the brightness is caused by increasing the temperature of some region on the surface of the star. The flare rise time is an indicator of heating this region on the surface. Therefore, the ratio of flare decay times to flare rise times, so the values of 3.5, must be a critical value between thermal or non-thermal processes.

As it is seen from the models of \citet{Gur88}, the differences between flare durations and flare amplitudes are seen between two flare types derived from observed flares in this study. The difference between amplitudes of slow and fast flares was given by Equation (22) in the paper of \citet{Gur88}. In the case of flare amplitude, the result obtained in this study is in agreement with this equation.

Providing that the value 3.5 is a limit ratio for flare types, fast flare rate is 63$\%$ of all 321 flares observed in this study, while slow flare rate is 37$\%$. It means that one of every three flares is a fast flare, while two of them are slow flares. This result diverges from what \citet{Gur88} stated. According to \citet{Gur88}, slow flares with low energies and low amplitudes are 95$\%$ of all flares. The remainder are fast flares. When looking individually over each star, the rate of flare types is changing from star to star. Detected flare number of \object{AD Leo} is 110 as it can be seen from Table 2. Slow flare rate of \object{AD Leo} flares is 78$\%$, while the rate of fast flares is 22$\%$. Detected flare number is 98 in observation of \object{EV Lac} and 40 in observation of V1054 Oph. Slow flare rates of both stars are 75$\%$, while fast flare rates are 25$\%$. \object{EQ Peg} flare number is 78. Slow flare rate of them is 63$\%$, and fast flare rate is 37$\%$.

In this study, one of the remarkable points is the correlation coefficients of linear fits. As it is seen in Table 4, the correlation coefficient is 0.732 for linear fit of slow flare type and 0.476 for fast flares. Although the correlation coefficient of the linear fit to the distribution of equivalent durations versus rise times is in an acceptable level for slow flares, it is relatively lower for fast flares. Regression calculations show that the best fits are linear for the distribution of equivalent durations versus rise times in logarithmic scales. The correlation coefficients of other fits are not higher than linear correlation coefficients. Especially, the correlation coefficient is lower for the fast flares due to the distribution of their equivalent durations. As it is seen from Figure 5, the equivalent durations of fast flares can take values in a wide range towards the longer riser times. This must be owing to the same reason of differences between \textit{y-intercept} values and the mean averages of equivalent durations of two flare types. As it is discussed above, while the slopes of the fits are nearly close to each other, there is a considerable difference between \textit{y-intercept} values and mean average of equivalent duration for two flare types. Consequently, all these deviations are seen in fast flares. The magnetic reconnection is dominant in this type of flares. A parameter in magnetic reconnection process causes some fast flares to be more powerful than the expected values. Eventually, some fast flares are more powerful than they are, while some of them are at expected energy levels. On the other hand, this parameter in magnetic reconnection process is not dominated in slow flare processes. And so, distribution of their equivalent durations is not scattered. This must be why the correlation coefficient of the fit is relatively higher for slow flares.

In this classification method, the complex flares are an exceptional case. These flares must be composed of some different flares. The complex flare should be separated into component flares before classification. If the fast and slow flares can be modelled, using these models, the complex flares can be decomposed into component flares. 

In conclusion, some parameters can be computed from flares observed in photoelectric photometry. And, if the behaviours between these parameters can be analysed by suitable methods, the flare types can be determined. In this study, we analysed the distributions of equivalent durations versus flare rise time by the statistical analysis method, \textit{t-Test}. Finally, it is seen that using the ratios of flare decay times to flare rise times, flares can be classified. Thus, flares are classified into two types as fast and slow flares. It is seen that there are considerable differences between these two types of the flares. The differences and the similarities between flare types are important to understand the flare processes. This gives new ideas to model white light flares of UV Ceti stars.

\section*{Acknowledgments} The authors acknowledge generous allotments of observing time at Ege University Observatory. We wish to thank both Dr. Hayal Boyac{\i}o\v{g}lu, who gave us important suggestions about statistical analyses, and Prof. Dr. M. Can Akan, who gave us valuable suggestions, which improved the language of the manuscript. We also thank the referee for useful comments which have contributed to improve the manuscript. We finally wish to thank the Ege University Research Found Council for supporting this work through grant Nr. 2005/FEN/051.

\clearpage

\begin{figure}
\epsscale{.80}
\plotone{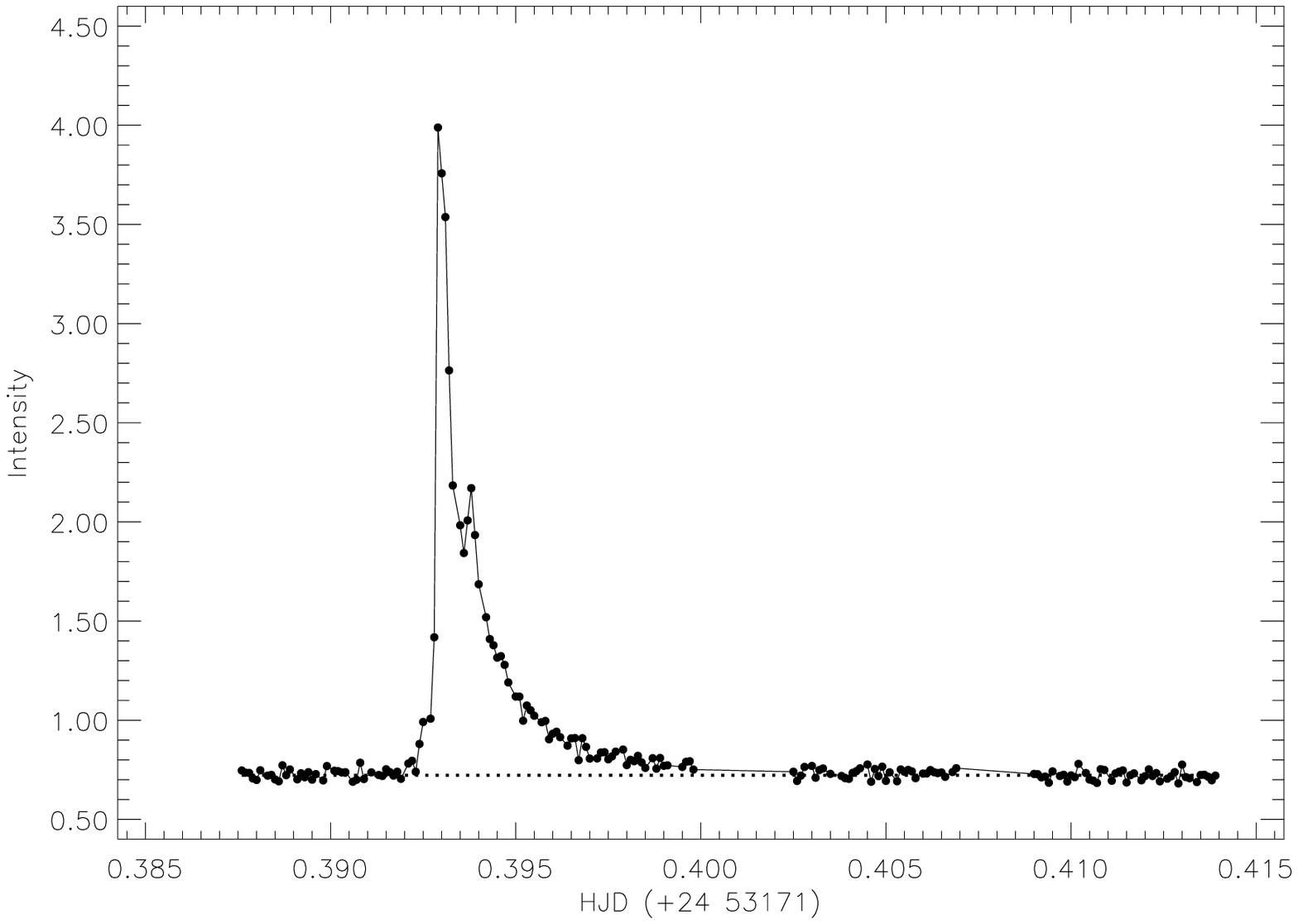}
\caption{A flare light curve sample for fast flares obtained from U band observation of V1054 Oph in 14 June 2004.\label{fig1}}
\end{figure}

\begin{figure}
\epsscale{.80}
\plotone{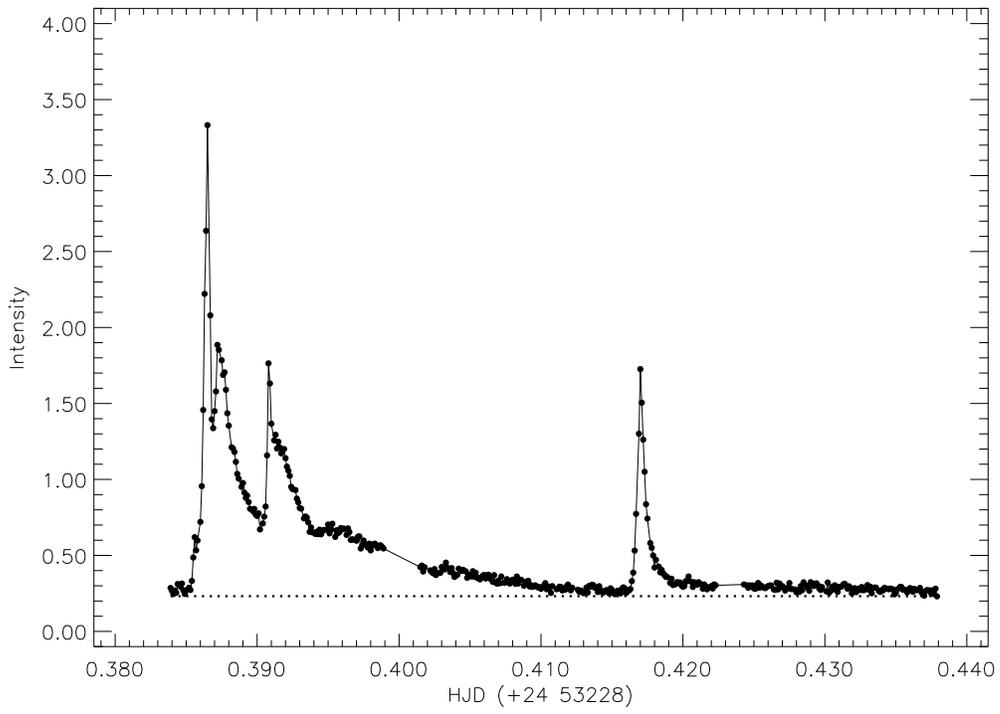}
\caption{A flare light curve sample for fast flares obtained from U band observation of  EV Lac in 10 August 2004.\label{fig2}}
\end{figure}

\begin{figure}
\epsscale{.80}
\plotone{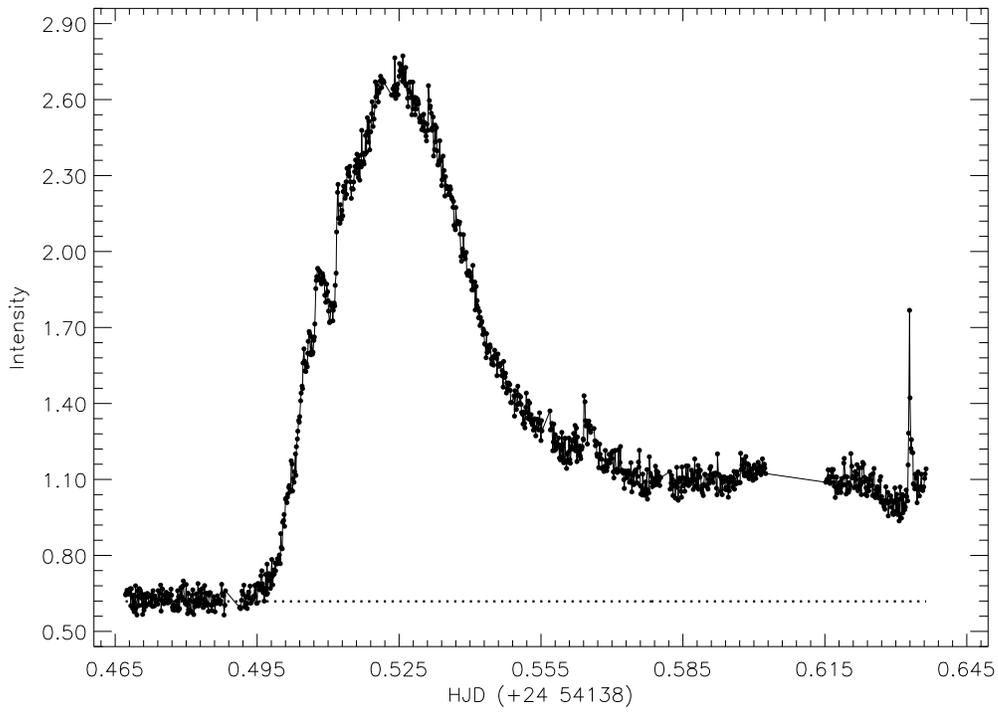}
\caption{A flare light curve sample for more powerful flares obtained from U band observation of AD Leo in 6 February 2007.\label{fig3}}
\end{figure}

\begin{figure}
\epsscale{.80}
\plotone{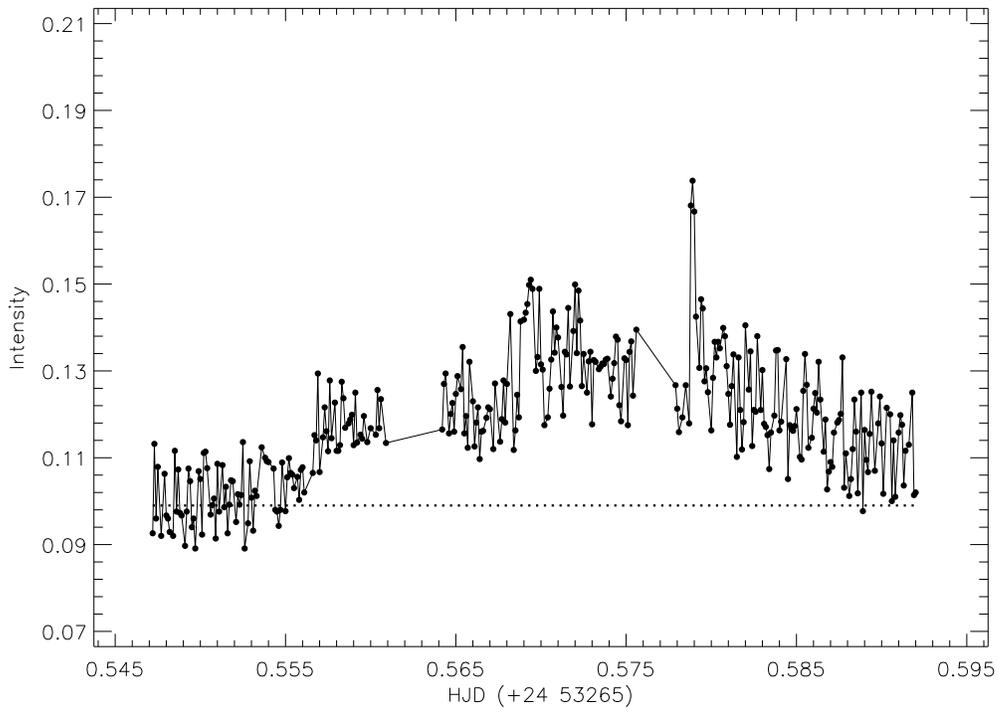}
\caption{A flare light curve sample for slow flares obtained from U band observation of EQ Peg in 16 September 2004.\label{fig4}}
\end{figure}

\begin{figure}
\epsscale{.80}
\plotone{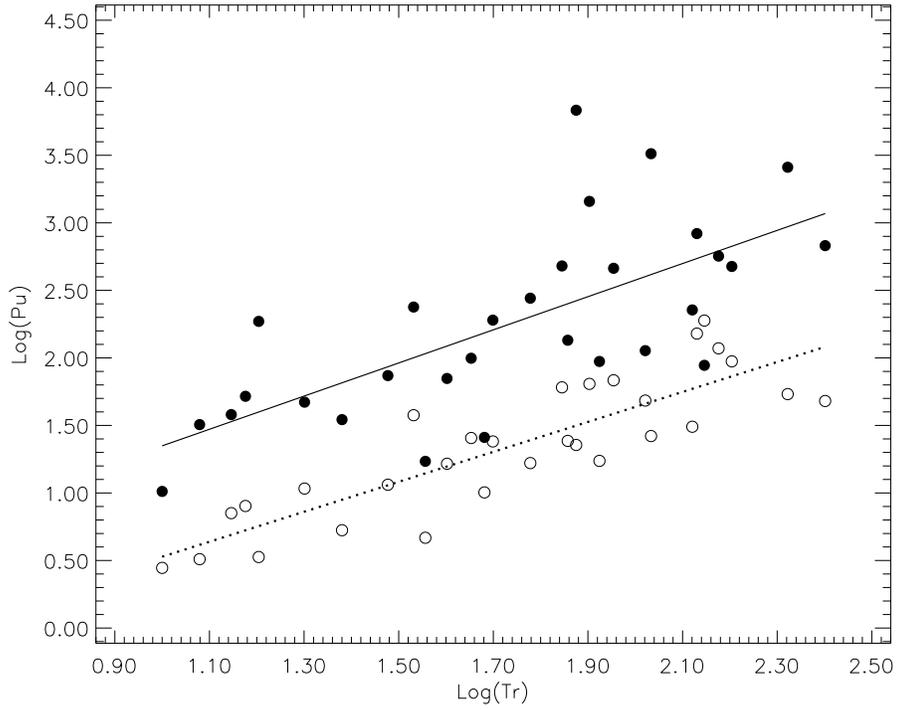}
\caption{The distributions for the mean averages of the equivalent durations ($Log (P_{u})$) versus flare rise times ($Log (T_{r})$) in logarithmic scale. In the figure, open circles represent slow flares, while filled circles show the fast flares. And the lines represent fits given equations (3) and (4).\label{fig5}}
\end{figure}

\begin{figure}
\epsscale{.80}
\plotone{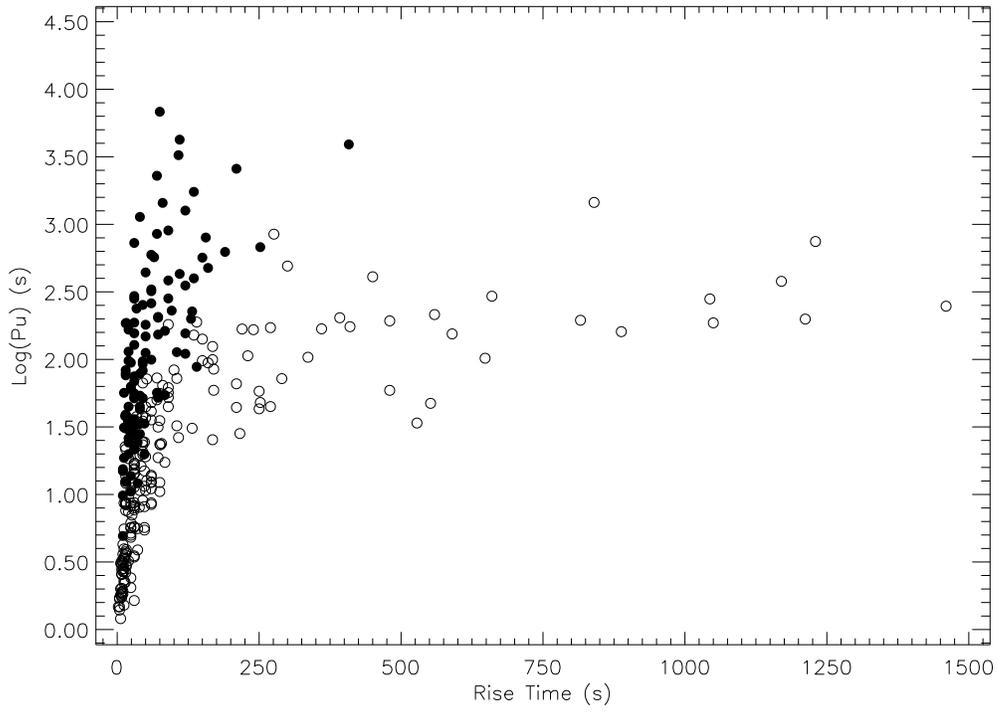}
\caption{The distributions of the equivalent durations ($Log(P_{u})$) in logarithmic scale versus flare rise times ($T_{r}$) for all 321 flares detected in observations of program stars. In the figure, open circles represent slow flares, while filled circles show the fast flares.\label{fig6}}
\end{figure}

\begin{figure}
\epsscale{.80}
\plotone{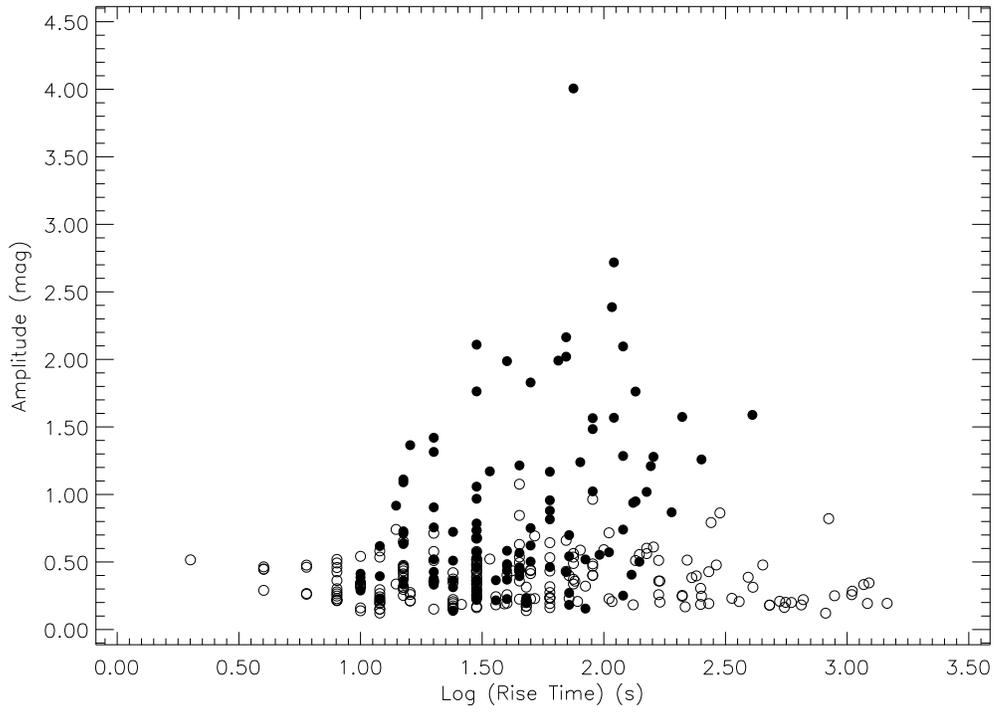}
\caption{The distributions of flare amplitudes versus flare rise times ($Log(T_{r})$) in logarithmic scale for all 321 flares detected in observations of program stars. In the figure, open circles represent slow flares, while filled circles show the fast flares.\label{fig7}}
\end{figure}

\clearpage

\begin{table}
\begin{center}
\caption{Basic parameters for the targets studied and their comparison (C1) and check (C2) stars.\label{tbl-1}}
\begin{tabular}{lrrrrr}
\\
\tableline\tableline
\textbf{Stars} & \textbf{V (mag)} & \textbf{B-V (mag)} & \textbf{Spectral Type} & \textbf{Distance (pc)} & \textbf{$LogL_{bol}$} \\
\tableline 
\textbf{AD Leo} & 9.388 & 1.498 & M3 & 4.90 & 31.87 \\
C1 = HD 89772 & 8.967 & 1.246 & K6-K7 & - & - \\
C2 = HD 89471 & 7.778 & 1.342 & K8 & - & - \\
\tableline 
\textbf{EV Lac} & 10.313 & 1.554 & M3 & 5.00 & 31.72 \\
C1 = HD 215576 & 9.227 & 1.197 & K6 & - & - \\
C2 = HD 215488 & 10.037 & 0.881 & K1-K2 & - & - \\
\tableline 
\textbf{EQ Peg} & 10.170 & 1.574 & M3-M4 & 6.58 & 31.42 \\
C1 = SAO 108666 & 9.598 & 0.745 & G8 & - & - \\
C2 = SAO 91312 & 9.050 & 1.040 & K3-K4 & - & - \\
\tableline
\textbf{V1054 Oph} & 8.996 & 1.552 & M3 & 5.70 & 31.93 \\
C1 = HD 152678 & 7.976 & 1.549 & M3 & - & - \\
C2 = SAO 141448 & 9.978 & 0.805 & K0 & - & - \\
\tableline
\end{tabular}
\end{center}
\end{table}

\begin{table}
\begin{center}
\caption{Observational reports of the each program star for each observing season.\label{tbl-2}}
\begin{tabular}{lcccccc}
\\
\tableline\tableline
\textbf{Stars} & \textbf{Year} & \textbf{HJD} & \textbf{Filter} & \textbf{Observation} & \textbf{Observation} & \textbf{U Filter} \\
& & \textbf{(+2400000)} & & \textbf{Number} & \textbf{Time (hour)} & \textbf{Flare Number} \\
\tableline 
\textbf{AD Leo} & 2005 & 53377 - 53514 & U & 12 & 36.35 & 39 \\
& 2006 & 53717 - 53831 & U & 15 & 37.48 & 54 \\
& 2007 & 54048 - 54248 & U & 8 & 20.80 & 17 \\
\tableline 
\textbf{EV Lac} & 2004 & 53197 - 53257 & U & 17 & 47.62 & 31 \\
& 2005 & 53554 - 53606 & U & 9 & 26.65 & 32 \\
& 2006 & 53940 - 53996 & U & 16 & 44.66 & 35 \\
\tableline
\textbf{EQ Peg} & 2004 & 53236 - 53335 & U & 13 & 64.42 & 38 \\
& 2005 & 53621 - 53686 & U & 10 & 35.84 & 35 \\
\tableline 
\textbf{V1054 Oph} & 2004 & 53136 - 53202 & U & 19 & 42.64 & 14 \\
& 2005 & 53502 - 53564 & U & 10 & 33.13 & 26 \\
\tableline
\end{tabular}
\end{center}
\end{table}

\rotate

\begin{deluxetable}{lrrrrrrrrr}
\tabletypesize{\scriptsize}
\tablecolumns{10}
\tablewidth{0pc}
\tablecaption{All the parameters were computed from observed flares. From the first column to the last, star name, the date of observation, HJD of flare maximum moment, flare total time (sec), decay time (sec), equivalent duration (sec), flare energy (erg), flare amplitude (mag) and flare type are given, respectively. And in the last column, it is specified whether the flare was used in the analyses, or not.} 
\\
\tablehead{
\colhead{Stars} & \colhead{Observation} & \colhead{HJD For} & \colhead{Total} & \colhead{Decay} & \colhead{Equivalent} & \colhead{Flare Energy} & \colhead{Amplitude} & \colhead{Flare} & \colhead{Used In} \\
\colhead{} & \colhead{Date} & \colhead{Maximum of Flare} & \colhead{Time (sec)} & \colhead{Time (sec)} & \colhead{Duration (sec)} & \colhead{(ergs)} & \colhead{In U Band (mag)} & \colhead{Type} & \colhead{Analyses}}
\\
\startdata
AD Leo 	&	 06.01.2005 	&	53377.50660	&	60	&	36	&	5.1690	&	6.8480E+30	&	0.348	&	Slow	&	Yes	\\
AD Leo 	&	 06.01.2005 	&	53377.59771	&	684	&	660	&	94.5297	&	1.2523E+32	&	0.723	&	Fast	&	Yes	\\
AD Leo 	&	 10.01.2005 	&	53381.51272	&	816	&	732	&	162.6476	&	2.1548E+32	&	0.519	&	Fast	&	Yes	\\
AD Leo 	&	 10.01.2005 	&	53381.52939	&	288	&	216	&	31.5306	&	4.1772E+31	&	0.399	&	Slow	&	Yes	\\
AD Leo 	&	 10.02.2005 	&	53412.49006	&	1308	&	1008	&	491.1727	&	6.5071E+32	&	0.863	&	Slow	&	No	\\
AD Leo 	&	 10.02.2005 	&	53412.52798	&	4164	&	3756	&	3907.4481	&	5.1767E+33	&	1.589	&	Fast	&	No	\\
AD Leo 	&	 10.02.2005 	&	53412.57464	&	1224	&	948	&	845.6318	&	1.1203E+33	&	0.792	&	Slow	&	No	\\
AD Leo 	&	 10.02.2005 	&	53412.58687	&	1764	&	1656	&	3250.0883	&	4.3058E+33	&	2.387	&	Fast	&	Yes	\\
AD Leo 	&	 11.02.2005 	&	53413.46661	&	36	&	24	&	3.0958	&	4.1014E+30	&	0.200	&	Slow	&	No	\\
AD Leo 	&	 11.02.2005 	&	53413.55759	&	384	&	360	&	60.6736	&	8.0382E+31	&	0.363	&	Fast	&	Yes	\\
AD Leo 	&	 12.03.2005 	&	53442.39045	&	720	&	680	&	78.4325	&	1.0391E+32	&	0.437	&	Fast	&	Yes	\\
AD Leo 	&	 14.03.2005 	&	53444.39896	&	432	&	300	&	30.9237	&	4.0968E+31	&	0.183	&	Slow	&	No	\\
AD Leo 	&	 14.03.2005 	&	53444.41327	&	1140	&	1068	&	152.7276	&	2.0234E+32	&	0.183	&	Fast	&	Yes	\\
AD Leo 	&	 14.03.2005 	&	53444.42674	&	588	&	504	&	54.4535	&	7.2141E+31	&	0.155	&	Fast	&	Yes	\\
AD Leo 	&	 14.03.2005 	&	53444.47105	&	1488	&	1428	&	320.0239	&	4.2397E+32	&	1.168	&	Fast	&	Yes	\\
AD Leo 	&	 14.03.2005 	&	53444.50563	&	1428	&	1356	&	206.4180	&	2.7347E+32	&	0.542	&	Fast	&	Yes	\\
AD Leo 	&	 14.03.2005 	&	53444.53049	&	240	&	72	&	25.4230	&	3.3681E+31	&	0.258	&	Slow	&	No	\\
AD Leo 	&	 14.03.2005 	&	53444.54716	&	576	&	324	&	47.9470	&	6.3521E+31	&	0.247	&	Slow	&	No	\\
AD Leo 	&	 14.03.2005 	&	53444.55396	&	84	&	60	&	10.7092	&	1.4188E+31	&	0.222	&	Slow	&	Yes	\\
AD Leo 	&	 14.03.2005 	&	53444.55507	&	84	&	48	&	5.5891	&	7.4046E+30	&	0.243	&	Slow	&	Yes	\\
AD Leo 	&	 14.03.2005 	&	53444.55632	&	132	&	72	&	8.6651	&	1.1480E+31	&	0.192	&	Slow	&	No	\\
AD Leo 	&	 14.03.2005 	&	53444.55966	&	24	&	12	&	8.6411	&	1.1448E+31	&	0.213	&	Slow	&	No	\\
AD Leo 	&	 14.03.2005 	&	53444.56021	&	60	&	36	&	8.6411	&	1.1448E+31	&	0.192	&	Slow	&	Yes	\\
AD Leo 	&	 14.03.2005 	&	53444.56313	&	204	&	156	&	24.4442	&	3.2384E+31	&	0.207	&	Slow	&	Yes	\\
AD Leo 	&	 16.03.2005 	&	53446.29405	&	2472	&	1812	&	293.6454	&	3.8903E+32	&	0.221	&	Slow	&	No	\\
AD Leo 	&	 16.03.2005 	&	53446.34849	&	48	&	24	&	2.4130	&	3.1968E+30	&	0.206	&	Slow	&	Yes	\\
AD Leo 	&	 16.03.2005 	&	53446.36113	&	456	&	408	&	33.4773	&	4.4351E+31	&	0.195	&	Fast	&	Yes	\\
AD Leo 	&	 16.03.2005 	&	53446.39558	&	2244	&	1032	&	198.8119	&	2.6339E+32	&	0.193	&	Slow	&	No	\\
AD Leo 	&	 16.03.2005 	&	53446.41252	&	852	&	756	&	229.8011	&	3.0444E+32	&	0.553	&	Fast	&	No	\\
AD Leo 	&	 16.03.2005 	&	53446.42960	&	756	&	588	&	99.5567	&	1.3189E+32	&	0.359	&	Slow	&	No	\\
AD Leo 	&	 16.03.2005 	&	53446.44822	&	132	&	72	&	12.2456	&	1.6223E+31	&	0.433	&	Slow	&	No	\\
AD Leo 	&	 09.04.2005 	&	53470.31255	&	1176	&	924	&	677.9344	&	8.9814E+32	&	1.259	&	Fast	&	No	\\
AD Leo 	&	 09.04.2005 	&	53470.33269	&	2088	&	1272	&	194.9401	&	2.5826E+32	&	0.121	&	Slow	&	No	\\
AD Leo 	&	 09.04.2005 	&	53470.36824	&	936	&	804	&	226.5383	&	3.0012E+32	&	0.938	&	Fast	&	No	\\
AD Leo 	&	 10.04.2005 	&	53471.30082	&	240	&	192	&	19.8520	&	2.6300E+31	&	0.234	&	Fast	&	Yes	\\
AD Leo 	&	 10.04.2005 	&	53471.37263	&	756	&	588	&	124.5185	&	1.6496E+32	&	0.512	&	Slow	&	No	\\
AD Leo 	&	 02.05.2005 	&	53493.31977	&	1284	&	1212	&	203.5007	&	2.6960E+32	&	0.699	&	Fast	&	Yes	\\
AD Leo 	&	 09.05.2005 	&	53500.32863	&	264	&	156	&	26.3800	&	3.4949E+31	&	0.207	&	Slow	&	No	\\
AD Leo 	&	 09.05.2005 	&	53500.35640	&	852	&	300	&	47.2850	&	6.2644E+31	&	0.164	&	Slow	&	No	\\
AD Leo 	&	 08.01.2006 	&	53744.50793	&	105	&	45	&	11.0325	&	1.4616E+31	&	0.236	&	Slow	&	No	\\
AD Leo 	&	 08.01.2006 	&	53744.55585	&	45	&	15	&	3.4390	&	4.5561E+30	&	0.162	&	Slow	&	No	\\
AD Leo 	&	 08.01.2006 	&	53744.64085	&	525	&	255	&	44.7429	&	5.9276E+31	&	0.192	&	Slow	&	No	\\
AD Leo 	&	 27.01.2006 	&	53763.62811	&	168	&	120	&	14.9402	&	1.9793E+31	&	0.171	&	Slow	&	Yes	\\
AD Leo 	&	 27.01.2006 	&	53763.63019	&	96	&	36	&	8.4330	&	1.1172E+31	&	0.162	&	Slow	&	No	\\
AD Leo 	&	 02.02.2006 	&	53769.52752	&	420	&	396	&	62.8336	&	8.3243E+31	&	0.510	&	Fast	&	Yes	\\
AD Leo 	&	 02.02.2006 	&	53769.55037	&	108	&	60	&	5.6936	&	7.5429E+30	&	0.140	&	Slow	&	Yes	\\
AD Leo 	&	 02.02.2006 	&	53769.61581	&	156	&	132	&	13.7184	&	1.8174E+31	&	0.138	&	Fast	&	Yes	\\
AD Leo 	&	 02.02.2006 	&	53769.61803	&	48	&	24	&	2.0431	&	2.7067E+30	&	0.161	&	Slow	&	Yes	\\
AD Leo 	&	 02.02.2006 	&	53769.62335	&	72	&	48	&	5.7062	&	7.5597E+30	&	0.160	&	Slow	&	Yes	\\
AD Leo 	&	 02.02.2006 	&	53769.62765	&	96	&	48	&	8.1257	&	1.0765E+31	&	0.173	&	Slow	&	Yes	\\
AD Leo 	&	 04.02.2006 	&	53771.46932	&	1188	&	1032	&	798.6289	&	1.0580E+33	&	1.210	&	Fast	&	No	\\
AD Leo 	&	 04.02.2006 	&	53771.49590	&	168	&	84	&	17.2768	&	2.2889E+31	&	0.321	&	Slow	&	Yes	\\
AD Leo 	&	 04.02.2006 	&	53771.50608	&	60	&	24	&	3.8830	&	5.1443E+30	&	0.183	&	Slow	&	Yes	\\
AD Leo 	&	 21.02.2006 	&	53788.42126	&	210	&	198	&	56.5353	&	7.4899E+31	&	0.618	&	Fast	&	No	\\
AD Leo 	&	 21.02.2006 	&	53788.42609	&	12	&	6	&	1.8036	&	2.3895E+30	&	0.261	&	Slow	&	No	\\
AD Leo 	&	 21.02.2006 	&	53788.49379	&	58	&	44	&	22.5737	&	2.9906E+31	&	0.741	&	Slow	&	Yes	\\
AD Leo 	&	 21.02.2006 	&	53788.49497	&	148	&	114	&	37.6468	&	4.9875E+31	&	0.522	&	Slow	&	Yes	\\
AD Leo 	&	 21.02.2006 	&	53788.49696	&	10	&	6	&	1.4739	&	1.9527E+30	&	0.290	&	Slow	&	No	\\
AD Leo 	&	 21.02.2006 	&	53788.53818	&	374	&	358	&	186.3927	&	2.4694E+32	&	1.365	&	Fast	&	Yes	\\
AD Leo 	&	 21.02.2006 	&	53788.54429	&	8	&	2	&	1.2042	&	1.5953E+30	&	0.267	&	Slow	&	No	\\
AD Leo 	&	 21.02.2006 	&	53788.54697	&	16	&	2	&	2.2268	&	2.9501E+30	&	0.337	&	Slow	&	Yes	\\
AD Leo 	&	 17.03.2006 	&	53812.29784	&	585	&	555	&	52.7018	&	6.9820E+31	&	0.294	&	Fast	&	Yes	\\
AD Leo 	&	 17.03.2006 	&	53812.32348	&	165	&	135	&	21.5461	&	2.8545E+31	&	0.230	&	Fast	&	Yes	\\
AD Leo 	&	 25.03.2006 	&	53820.28531	&	12	&	4	&	3.0728	&	4.0708E+30	&	0.363	&	Slow	&	No	\\
AD Leo 	&	 25.03.2006 	&	53820.31442	&	138	&	62	&	23.2816	&	3.0844E+31	&	0.358	&	Slow	&	No	\\
AD Leo 	&	 25.03.2006 	&	53820.31970	&	14	&	2	&	2.8248	&	3.7423E+30	&	0.582	&	Slow	&	No	\\
AD Leo 	&	 25.03.2006 	&	53820.40075	&	256	&	222	&	237.9794	&	3.1528E+32	&	1.171	&	Fast	&	Yes	\\
AD Leo 	&	 25.03.2006 	&	53820.40674	&	90	&	48	&	16.2012	&	2.1464E+31	&	0.435	&	Slow	&	No	\\
AD Leo 	&	 25.03.2006 	&	53820.40742	&	20	&	10	&	3.1643	&	4.1921E+30	&	0.542	&	Slow	&	No	\\
AD Leo 	&	 25.03.2006 	&	53820.41637	&	8	&	4	&	1.7195	&	2.2781E+30	&	0.448	&	Slow	&	No	\\
AD Leo 	&	 25.03.2006 	&	53820.41686	&	12	&	6	&	3.0961	&	4.1018E+30	&	0.477	&	Slow	&	No	\\
AD Leo 	&	 01.04.2006 	&	53827.27390	&	10	&	6	&	1.3930	&	1.8455E+30	&	0.464	&	Slow	&	No	\\
AD Leo 	&	 01.04.2006 	&	53827.29613	&	12	&	4	&	2.5516	&	3.3804E+30	&	0.493	&	Slow	&	No	\\
AD Leo 	&	 01.04.2006 	&	53827.30144	&	10	&	4	&	2.0011	&	2.6511E+30	&	0.461	&	Slow	&	No	\\
AD Leo 	&	 01.04.2006 	&	53827.31787	&	10	&	2	&	1.8332	&	2.4286E+30	&	0.519	&	Slow	&	No	\\
AD Leo 	&	 01.04.2006 	&	53827.37920	&	12	&	4	&	1.8984	&	2.5150E+30	&	0.427	&	Slow	&	No	\\
AD Leo 	&	 05.04.2006 	&	53831.39559	&	66	&	46	&	34.7529	&	4.6041E+31	&	0.713	&	Slow	&	Yes	\\
AD Leo 	&	 05.04.2006 	&	53831.39755	&	10	&	6	&	1.6828	&	2.2295E+30	&	0.446	&	Slow	&	No	\\
AD Leo 	&	 05.04.2006 	&	53831.40919	&	76	&	62	&	38.0426	&	5.0400E+31	&	0.917	&	Fast	&	No	\\
AD Leo 	&	 05.04.2006 	&	53831.41058	&	16	&	8	&	2.7778	&	3.6801E+30	&	0.459	&	Slow	&	No	\\
AD Leo 	&	 05.04.2006 	&	53831.41937	&	14	&	2	&	3.3755	&	4.4720E+30	&	0.536	&	Slow	&	No	\\
AD Leo 	&	 01.12.2006 	&	54071.52083	&	78	&	39	&	8.0437	&	1.0657E+31	&	0.190	&	Slow	&	No	\\
AD Leo 	&	 01.12.2006 	&	54071.52940	&	156	&	104	&	12.6557	&	1.6767E+31	&	0.229	&	Slow	&	No	\\
AD Leo 	&	 01.12.2006 	&	54071.53617	&	208	&	130	&	23.8354	&	3.1578E+31	&	0.208	&	Slow	&	No	\\
AD Leo 	&	 01.12.2006 	&	54071.54174	&	52	&	26	&	7.0330	&	9.3175E+30	&	0.186	&	Slow	&	No	\\
AD Leo 	&	 01.12.2006 	&	54071.59966	&	2258	&	1699	&	214.7257	&	2.8447E+32	&	0.201	&	Slow	&	No	\\
AD Leo 	&	 15.12.2006 	&	54085.63076	&	20	&	10	&	1.8367	&	2.4332E+30	&	0.139	&	Slow	&	No	\\
AD Leo 	&	 15.12.2006 	&	54085.63572	&	30	&	20	&	1.9239	&	2.5488E+30	&	0.160	&	Slow	&	No	\\
AD Leo 	&	 23.12.2006 	&	54093.60562	&	30	&	10	&	3.2230	&	4.2698E+30	&	0.151	&	Slow	&	Yes	\\
AD Leo 	&	 23.12.2006 	&	54093.60712	&	40	&	10	&	3.5123	&	4.6532E+30	&	0.170	&	Slow	&	No	\\
AD Leo 	&	 23.12.2006 	&	54093.61349	&	150	&	110	&	10.6923	&	1.4165E+31	&	0.195	&	Slow	&	No	\\
AD Leo 	&	 23.12.2006 	&	54093.63486	&	650	&	400	&	58.1135	&	7.6990E+31	&	0.186	&	Slow	&	No	\\
AD Leo 	&	 23.12.2006 	&	54093.64631	&	1676	&	1086	&	154.4312	&	2.0459E+32	&	0.200	&	Slow	&	No	\\
AD Leo 	&	 21.01.2007 	&	54122.47422	&	16	&	8	&	1.8714	&	2.4792E+30	&	0.229	&	Slow	&	No	\\
AD Leo 	&	 21.01.2007 	&	54122.48635	&	32	&	16	&	3.9082	&	5.1777E+30	&	0.274	&	Slow	&	Yes	\\
AD Leo 	&	 21.01.2007 	&	54122.48876	&	16	&	8	&	2.0181	&	2.6736E+30	&	0.302	&	Slow	&	No	\\
AD Leo 	&	 21.01.2007 	&	54122.49107	&	32	&	16	&	3.6712	&	4.8637E+30	&	0.213	&	Slow	&	Yes	\\
AD Leo 	&	 21.01.2007 	&	54122.49573	&	24	&	16	&	2.9831	&	3.9520E+30	&	0.248	&	Slow	&	No	\\
AD Leo 	&	 21.01.2007 	&	54122.49693	&	16	&	8	&	1.7260	&	2.2867E+30	&	0.264	&	Slow	&	No	\\
AD Leo 	&	 21.01.2007 	&	54122.49712	&	16	&	8	&	1.7700	&	2.3449E+30	&	0.215	&	Slow	&	No	\\
AD Leo 	&	 21.01.2007 	&	54122.49860	&	16	&	8	&	1.8787	&	2.4889E+30	&	0.214	&	Slow	&	No	\\
AD Leo 	&	 21.01.2007 	&	54122.50156	&	24	&	8	&	2.9281	&	3.8792E+30	&	0.259	&	Slow	&	Yes	\\
AD Leo 	&	 21.01.2007 	&	54122.50582	&	40	&	24	&	3.0146	&	3.9938E+30	&	0.211	&	Slow	&	Yes	\\
AD Leo 	&	 21.01.2007 	&	54122.50628	&	24	&	16	&	2.5792	&	3.4170E+30	&	0.217	&	Slow	&	No	\\
AD Leo 	&	 21.01.2007 	&	54122.53384	&	24	&	16	&	3.2242	&	4.2715E+30	&	0.279	&	Slow	&	No	\\
AD Leo 	&	 08.03.2007 	&	54168.46811	&	60	&	30	&	8.2226	&	1.0893E+31	&	0.285	&	Slow	&	No	\\
AD Leo 	&	 08.03.2007 	&	54168.46933	&	90	&	60	&	14.2830	&	1.8922E+31	&	0.232	&	Slow	&	No	\\
AD Leo 	&	 16.03.2007 	&	54176.37237	&	84	&	60	&	4.8172	&	6.3820E+30	&	0.157	&	Slow	&	Yes	\\
AD Leo 	&	 16.03.2007 	&	54176.39611	&	192	&	120	&	18.7237	&	2.4805E+31	&	0.230	&	Slow	&	Yes	\\
AD Leo 	&	 16.03.2007 	&	54176.45538	&	372	&	156	&	28.2549	&	3.7433E+31	&	0.168	&	Slow	&	No	\\
EQ Peg 	&	 18.08.2004 	&	53236.43176	&	290	&	250	&	44.8869	&	4.4695E+31	&	0.369	&	Fast	&	Yes	\\
EQ Peg 	&	 18.08.2004 	&	53236.48106	&	210	&	170	&	28.0058	&	2.7886E+31	&	0.226	&	Fast	&	Yes	\\
EQ Peg 	&	 18.08.2004 	&	53236.50398	&	220	&	180	&	42.7730	&	4.2590E+31	&	0.473	&	Fast	&	Yes	\\
EQ Peg 	&	 18.08.2004 	&	53236.56139	&	1060	&	900	&	475.3540	&	4.7332E+32	&	1.279	&	Fast	&	No	\\
EQ Peg 	&	 19.08.2004 	&	53237.38157	&	490	&	440	&	111.7547	&	1.1128E+32	&	0.503	&	Fast	&	Yes	\\
EQ Peg 	&	 19.08.2004 	&	53237.40715	&	150	&	110	&	34.0142	&	3.3869E+31	&	0.258	&	Slow	&	No	\\
EQ Peg 	&	 19.08.2004 	&	53237.41224	&	340	&	310	&	68.8944	&	6.8600E+31	&	0.464	&	Fast	&	Yes	\\
EQ Peg 	&	 19.08.2004 	&	53237.48805	&	230	&	200	&	51.5306	&	5.1310E+31	&	0.734	&	Fast	&	Yes	\\
EQ Peg 	&	 19.08.2004 	&	53237.49291	&	180	&	160	&	30.2882	&	3.0159E+31	&	0.343	&	Fast	&	Yes	\\
EQ Peg 	&	 19.08.2004 	&	53237.53122	&	610	&	380	&	106.3842	&	1.0593E+32	&	0.384	&	Slow	&	No	\\
EQ Peg 	&	 22.08.2004 	&	53240.36934	&	570	&	510	&	260.2007	&	2.5909E+32	&	0.880	&	Fast	&	Yes	\\
EQ Peg 	&	 08.09.2004 	&	53257.36799	&	340	&	270	&	56.6293	&	5.6387E+31	&	0.427	&	Fast	&	No	\\
EQ Peg 	&	 08.09.2004 	&	53257.38223	&	430	&	140	&	72.0350	&	7.1727E+31	&	0.478	&	Slow	&	No	\\
EQ Peg 	&	 08.09.2004 	&	53257.45202	&	1170	&	1040	&	199.7623	&	1.9891E+32	&	0.406	&	Fast	&	No	\\
EQ Peg 	&	 08.09.2004 	&	53257.51000	&	120	&	110	&	14.8846	&	1.4821E+31	&	0.413	&	Fast	&	No	\\
EQ Peg 	&	 08.09.2004 	&	53257.51845	&	60	&	30	&	11.3037	&	1.1255E+31	&	0.520	&	Slow	&	No	\\
EQ Peg 	&	 09.09.2004 	&	53258.36870	&	1840	&	1390	&	409.0358	&	4.0729E+32	&	0.478	&	Slow	&	No	\\
EQ Peg 	&	 09.09.2004 	&	53258.40585	&	190	&	160	&	32.7460	&	3.2606E+31	&	0.574	&	Fast	&	Yes	\\
EQ Peg 	&	 09.09.2004 	&	53258.51094	&	200	&	160	&	33.3344	&	3.3192E+31	&	0.484	&	Fast	&	Yes	\\
EQ Peg 	&	 12.09.2004 	&	53261.39106	&	260	&	220	&	53.8905	&	5.3660E+31	&	0.583	&	Fast	&	Yes	\\
EQ Peg 	&	 12.09.2004 	&	53261.46988	&	130	&	110	&	19.8878	&	1.9803E+31	&	0.332	&	Fast	&	Yes	\\
EQ Peg 	&	 12.09.2004 	&	53261.55529	&	150	&	130	&	35.5859	&	3.5434E+31	&	0.756	&	Fast	&	Yes	\\
EQ Peg 	&	 12.09.2004 	&	53261.59048	&	210	&	180	&	74.9052	&	7.4585E+31	&	0.785	&	Fast	&	Yes	\\
EQ Peg 	&	 14.09.2004 	&	53263.34072	&	1070	&	950	&	110.0519	&	1.0958E+32	&	0.251	&	Fast	&	No	\\
EQ Peg 	&	 14.09.2004 	&	53263.36815	&	190	&	130	&	41.2794	&	4.1103E+31	&	0.532	&	Slow	&	No	\\
EQ Peg 	&	 14.09.2004 	&	53263.37752	&	1620	&	1430	&	625.0200	&	6.2235E+32	&	0.868	&	Fast	&	No	\\
EQ Peg 	&	 14.09.2004 	&	53263.40484	&	200	&	150	&	44.5257	&	4.4335E+31	&	0.414	&	Slow	&	Yes	\\
EQ Peg 	&	 14.09.2004 	&	53263.47371	&	50	&	40	&	4.9279	&	4.9068E+30	&	0.289	&	Fast	&	No	\\
EQ Peg 	&	 14.09.2004 	&	53263.48725	&	70	&	50	&	9.4754	&	9.4349E+30	&	0.357	&	Slow	&	Yes	\\
EQ Peg 	&	 14.09.2004 	&	53263.54697	&	60	&	30	&	5.7809	&	5.7562E+30	&	0.361	&	Slow	&	No	\\
EQ Peg 	&	 15.09.2004 	&	53264.46607	&	330	&	120	&	44.1024	&	4.3914E+31	&	0.246	&	Slow	&	No	\\
EQ Peg 	&	 15.09.2004 	&	53264.50102	&	580	&	550	&	294.4328	&	2.9317E+32	&	2.109	&	Fast	&	Yes	\\
EQ Peg 	&	 15.09.2004 	&	53264.51167	&	510	&	490	&	166.0980	&	1.6539E+32	&	1.420	&	Fast	&	Yes	\\
EQ Peg 	&	 15.09.2004 	&	53264.56595	&	1440	&	390	&	186.4048	&	1.8561E+32	&	0.284	&	Slow	&	No	\\
EQ Peg 	&	 16.09.2004 	&	53265.39674	&	910	&	860	&	180.5574	&	1.7979E+32	&	0.622	&	Fast	&	Yes	\\
EQ Peg 	&	 16.09.2004 	&	53265.51329	&	420	&	250	&	84.7088	&	8.4347E+31	&	0.203	&	Slow	&	No	\\
EQ Peg 	&	 16.09.2004 	&	53265.54003	&	410	&	350	&	99.4698	&	9.9044E+31	&	0.463	&	Fast	&	Yes	\\
EQ Peg 	&	 16.09.2004 	&	53265.56942	&	3180	&	1950	&	746.5378	&	7.4335E+32	&	0.345	&	Slow	&	No	\\
EQ Peg 	&	 07.09.2005 	&	53621.52654	&	150	&	105	&	34.9365	&	3.4787E+31	&	0.630	&	Slow	&	No	\\
EQ Peg 	&	 08.09.2005 	&	53622.45159	&	285	&	270	&	78.7774	&	7.8441E+31	&	0.711	&	Fast	&	Yes	\\
EQ Peg 	&	 08.09.2005 	&	53622.46218	&	540	&	510	&	156.1378	&	1.5547E+32	&	0.968	&	Fast	&	Yes	\\
EQ Peg 	&	 08.09.2005 	&	53622.48087	&	120	&	105	&	39.0373	&	3.8870E+31	&	0.640	&	Fast	&	Yes	\\
EQ Peg 	&	 08.09.2005 	&	53622.49927	&	75	&	60	&	12.4900	&	1.2437E+31	&	0.362	&	Fast	&	Yes	\\
EQ Peg 	&	 12.09.2005 	&	53626.33898	&	1839	&	1764	&	6818.1425	&	6.7890E+33	&	4.006	&	Fast	&	Yes	\\
EQ Peg 	&	 12.09.2005 	&	53626.36062	&	75	&	45	&	8.6906	&	8.6534E+30	&	0.261	&	Slow	&	No	\\
EQ Peg 	&	 12.09.2005 	&	53626.36218	&	105	&	75	&	15.2364	&	1.5171E+31	&	0.338	&	Slow	&	No	\\
EQ Peg 	&	 12.09.2005 	&	53626.36409	&	105	&	30	&	10.5085	&	1.0464E+31	&	0.337	&	Slow	&	Yes	\\
EQ Peg 	&	 12.09.2005 	&	53626.36582	&	922	&	802	&	352.2107	&	3.5070E+32	&	1.285	&	Fast	&	No	\\
EQ Peg 	&	 12.09.2005 	&	53626.40874	&	225	&	180	&	51.6810	&	5.1460E+31	&	0.567	&	Fast	&	Yes	\\
EQ Peg 	&	 12.09.2005 	&	53626.50223	&	90	&	45	&	24.2883	&	2.4184E+31	&	0.440	&	Slow	&	No	\\
EQ Peg 	&	 12.09.2005 	&	53626.56201	&	315	&	225	&	181.2224	&	1.8045E+32	&	0.964	&	Slow	&	Yes	\\
EQ Peg 	&	 12.09.2005 	&	53626.56478	&	135	&	120	&	83.2370	&	8.2881E+31	&	1.090	&	Fast	&	Yes	\\
EQ Peg 	&	 12.09.2005 	&	53626.56635	&	60	&	45	&	12.1029	&	1.2051E+31	&	0.252	&	Slow	&	Yes	\\
EQ Peg 	&	 12.09.2005 	&	53626.56721	&	180	&	150	&	27.7286	&	2.7610E+31	&	0.282	&	Fast	&	Yes	\\
EQ Peg 	&	 12.09.2005 	&	53626.56930	&	60	&	30	&	5.7164	&	5.6920E+30	&	0.248	&	Slow	&	No	\\
EQ Peg 	&	 12.09.2005 	&	53626.58426	&	165	&	135	&	31.8950	&	3.1759E+31	&	0.676	&	Fast	&	Yes	\\
EQ Peg 	&	 27.09.2005 	&	53641.34719	&	135	&	120	&	76.2420	&	7.5916E+31	&	1.111	&	Fast	&	Yes	\\
EQ Peg 	&	 27.09.2005 	&	53641.36437	&	90	&	45	&	9.0307	&	8.9921E+30	&	0.225	&	Slow	&	No	\\
EQ Peg 	&	 27.09.2005 	&	53641.39000	&	150	&	120	&	35.8623	&	3.5709E+31	&	0.535	&	Fast	&	Yes	\\
EQ Peg 	&	 27.09.2005 	&	53641.42135	&	1545	&	1410	&	1739.8937	&	1.7325E+33	&	1.762	&	Fast	&	No	\\
EQ Peg 	&	 27.09.2005 	&	53641.47316	&	977	&	585	&	203.3585	&	2.0249E+32	&	0.387	&	Slow	&	No	\\
EQ Peg 	&	 03.10.2005 	&	53647.34180	&	180	&	135	&	66.8966	&	6.6611E+31	&	1.076	&	Slow	&	No	\\
EQ Peg 	&	 03.10.2005 	&	53647.40952	&	255	&	180	&	56.0173	&	5.5778E+31	&	0.536	&	Slow	&	Yes	\\
EQ Peg 	&	 28.10.2005 	&	53672.27432	&	2764	&	1594	&	378.4947	&	3.7688E+32	&	0.333	&	Slow	&	No	\\
EQ Peg 	&	 28.10.2005 	&	53672.30797	&	210	&	120	&	52.2187	&	5.1995E+31	&	0.402	&	Slow	&	Yes	\\
EQ Peg 	&	 28.10.2005 	&	53672.33656	&	225	&	195	&	55.9946	&	5.5755E+31	&	0.682	&	Fast	&	Yes	\\
EQ Peg 	&	 29.10.2005 	&	53673.34763	&	570	&	450	&	155.7805	&	1.5511E+32	&	0.740	&	Fast	&	No	\\
EQ Peg 	&	 29.10.2005 	&	53673.35896	&	240	&	135	&	72.3348	&	7.2025E+31	&	0.716	&	Slow	&	No	\\
EQ Peg 	&	 29.10.2005 	&	53673.36104	&	330	&	285	&	96.7821	&	9.6368E+31	&	0.461	&	Fast	&	Yes	\\
EQ Peg 	&	 29.10.2005 	&	53673.38314	&	738	&	468	&	171.8148	&	1.7108E+32	&	0.428	&	Slow	&	No	\\
EQ Peg 	&	 11.11.2005 	&	53686.29364	&	540	&	405	&	151.3632	&	1.5072E+32	&	0.511	&	Slow	&	No	\\
EQ Peg 	&	 11.11.2005 	&	53686.30076	&	795	&	645	&	566.9639	&	5.6454E+32	&	1.019	&	Fast	&	No	\\
EQ Peg 	&	 11.11.2005 	&	53686.37753	&	270	&	120	&	97.9336	&	9.7515E+31	&	0.562	&	Slow	&	No	\\
EV Lac 	&	 11.07.2004 	&	53198.47514	&	260	&	190	&	50.2493	&	2.3691E+31	&	0.431	&	Slow	&	No	\\
EV Lac 	&	 11.07.2004 	&	53198.48325	&	590	&	560	&	128.0124	&	6.0353E+31	&	0.735	&	Fast	&	Yes	\\
EV Lac 	&	 17.07.2004 	&	53204.51795	&	1230	&	820	&	174.6712	&	8.2351E+31	&	0.313	&	Slow	&	No	\\
EV Lac 	&	 20.07.2004 	&	53207.49970	&	310	&	230	&	64.2165	&	3.0276E+31	&	0.588	&	Slow	&	No	\\
EV Lac 	&	 20.07.2004 	&	53207.51648	&	510	&	490	&	114.2672	&	5.3873E+31	&	0.905	&	Fast	&	Yes	\\
EV Lac 	&	 24.07.2004 	&	53211.54445	&	830	&	590	&	165.8313	&	7.8184E+31	&	0.397	&	Slow	&	No	\\
EV Lac 	&	 25.07.2004 	&	53212.48570	&	490	&	340	&	141.3253	&	6.6630E+31	&	0.602	&	Slow	&	No	\\
EV Lac 	&	 25.07.2004 	&	53212.51625	&	380	&	160	&	167.7941	&	7.9109E+31	&	0.514	&	Slow	&	No	\\
EV Lac 	&	 26.07.2004 	&	53213.48702	&	80	&	50	&	17.0178	&	8.0233E+30	&	0.676	&	Slow	&	No	\\
EV Lac 	&	 26.07.2004 	&	53213.52209	&	100	&	90	&	15.4025	&	7.2618E+30	&	0.385	&	Fast	&	No	\\
EV Lac 	&	 28.07.2004 	&	53215.48409	&	180	&	160	&	97.5996	&	4.6015E+31	&	1.315	&	Fast	&	Yes	\\
EV Lac 	&	 28.07.2004 	&	53215.49960	&	170	&	120	&	36.9820	&	1.7436E+31	&	0.436	&	Slow	&	Yes	\\
EV Lac 	&	 07.08.2004 	&	53225.42498	&	350	&	190	&	94.2676	&	4.4444E+31	&	0.611	&	Slow	&	No	\\
EV Lac 	&	 07.08.2004 	&	53225.46827	&	140	&	80	&	47.9538	&	2.2609E+31	&	0.643	&	Slow	&	No	\\
EV Lac 	&	 08.08.2004 	&	53226.46483	&	140	&	120	&	26.0719	&	1.2292E+31	&	0.377	&	Fast	&	Yes	\\
EV Lac 	&	 08.08.2004 	&	53226.50951	&	2482	&	2422	&	594.8340	&	2.8044E+32	&	0.816	&	Fast	&	Yes	\\
EV Lac 	&	 09.08.2004 	&	53227.38165	&	940	&	910	&	728.2759	&	3.4336E+32	&	1.763	&	Fast	&	Yes	\\
EV Lac 	&	 09.08.2004 	&	53227.42505	&	630	&	490	&	188.9706	&	8.9093E+31	&	0.555	&	Slow	&	No	\\
EV Lac 	&	 09.08.2004 	&	53227.43928	&	360	&	260	&	83.5153	&	3.9375E+31	&	0.591	&	Slow	&	No	\\
EV Lac 	&	 09.08.2004 	&	53227.44287	&	240	&	220	&	44.6927	&	2.1071E+31	&	0.427	&	Fast	&	Yes	\\
EV Lac 	&	 09.08.2004 	&	53227.46382	&	1000	&	960	&	1134.4908	&	5.3487E+32	&	1.987	&	Fast	&	Yes	\\
EV Lac 	&	 09.08.2004 	&	53227.49692	&	90	&	70	&	18.4229	&	8.6857E+30	&	0.511	&	Slow	&	Yes	\\
EV Lac 	&	 09.08.2004 	&	53227.55063	&	70	&	60	&	9.8307	&	4.6348E+30	&	0.353	&	Fast	&	No	\\
EV Lac 	&	 10.08.2004 	&	53228.37056	&	90	&	50	&	19.5669	&	9.2251E+30	&	0.404	&	Slow	&	No	\\
EV Lac 	&	 10.08.2004 	&	53228.38654	&	2200	&	2090	&	4235.9481	&	1.9971E+33	&	2.718	&	Fast	&	No	\\
EV Lac 	&	 10.08.2004 	&	53228.41698	&	1800	&	1730	&	849.9341	&	4.0071E+32	&	2.020	&	Fast	&	No	\\
EV Lac 	&	 10.08.2004 	&	53228.52716	&	50	&	20	&	8.2244	&	3.8775E+30	&	0.437	&	Slow	&	No	\\
EV Lac 	&	 10.08.2004 	&	53228.55471	&	120	&	100	&	24.2433	&	1.1430E+31	&	0.520	&	Fast	&	Yes	\\
EV Lac 	&	 14.08.2004 	&	53232.36686	&	80	&	50	&	14.2793	&	6.7322E+30	&	0.433	&	Slow	&	No	\\
EV Lac 	&	 14.08.2004 	&	53232.37785	&	290	&	220	&	72.8660	&	3.4354E+31	&	0.660	&	Slow	&	No	\\
EV Lac 	&	 14.08.2004 	&	53232.39359	&	520	&	470	&	147.9080	&	6.9733E+31	&	0.751	&	Fast	&	Yes	\\
EV Lac 	&	 05.07.2005 	&	53557.52397	&	168	&	116	&	71.6667	&	3.3788E+31	&	0.693	&	Slow	&	No	\\
EV Lac 	&	 05.07.2005 	&	53557.53656	&	4	&	2	&	1.4724	&	6.9419E+29	&	0.516	&	Slow	&	No	\\
EV Lac 	&	 26.07.2005 	&	53578.50257	&	600	&	585	&	184.9502	&	8.7197E+31	&	0.727	&	Fast	&	Yes	\\
EV Lac 	&	 26.07.2005 	&	53578.51594	&	105	&	75	&	15.5441	&	7.3285E+30	&	0.381	&	Slow	&	No	\\
EV Lac 	&	 26.07.2005 	&	53578.52566	&	30	&	15	&	3.7640	&	1.7746E+30	&	0.381	&	Slow	&	Yes	\\
EV Lac 	&	 26.07.2005 	&	53578.54667	&	120	&	45	&	12.2619	&	5.7810E+30	&	0.372	&	Slow	&	Yes	\\
EV Lac 	&	 03.08.2005 	&	53586.50688	&	570	&	480	&	384.1497	&	1.8111E+32	&	1.565	&	Fast	&	Yes	\\
EV Lac 	&	 03.08.2005 	&	53586.51348	&	30	&	15	&	2.6479	&	1.2484E+30	&	0.313	&	Slow	&	Yes	\\
EV Lac 	&	 03.08.2005 	&	53586.52856	&	180	&	135	&	38.1427	&	1.7983E+31	&	0.437	&	Slow	&	No	\\
EV Lac 	&	 14.08.2005 	&	53597.33755	&	135	&	105	&	41.7339	&	1.9676E+31	&	0.361	&	Slow	&	No	\\
EV Lac 	&	 14.08.2005 	&	53597.40029	&	60	&	45	&	8.4471	&	3.9825E+30	&	0.297	&	Slow	&	Yes	\\
EV Lac 	&	 14.08.2005 	&	53597.47199	&	105	&	60	&	23.3056	&	1.0988E+31	&	0.359	&	Slow	&	No	\\
EV Lac 	&	 14.08.2005 	&	53597.47615	&	210	&	120	&	44.6968	&	2.1073E+31	&	0.401	&	Slow	&	Yes	\\
EV Lac 	&	 14.08.2005 	&	53597.47806	&	105	&	75	&	35.7253	&	1.6843E+31	&	0.304	&	Slow	&	No	\\
EV Lac 	&	 14.08.2005 	&	53597.48015	&	105	&	75	&	30.4018	&	1.4333E+31	&	0.493	&	Slow	&	No	\\
EV Lac 	&	 14.08.2005 	&	53597.48223	&	165	&	60	&	32.2519	&	1.5206E+31	&	0.228	&	Slow	&	No	\\
EV Lac 	&	 15.08.2005 	&	53598.42911	&	390	&	345	&	92.8472	&	4.3774E+31	&	0.435	&	Fast	&	Yes	\\
EV Lac 	&	 15.08.2005 	&	53598.43380	&	195	&	165	&	33.7418	&	1.5908E+31	&	0.219	&	Fast	&	Yes	\\
EV Lac 	&	 15.08.2005 	&	53598.46517	&	1005	&	870	&	398.3372	&	1.8780E+32	&	0.949	&	Fast	&	No	\\
EV Lac 	&	 15.08.2005 	&	53598.50008	&	810	&	720	&	282.6663	&	1.3327E+32	&	1.023	&	Fast	&	Yes	\\
EV Lac 	&	 15.08.2005 	&	53598.52741	&	1365	&	1245	&	1264.2000	&	5.9603E+32	&	2.096	&	Fast	&	No	\\
EV Lac 	&	 22.08.2005 	&	53605.39060	&	60	&	45	&	21.6520	&	1.0208E+31	&	0.467	&	Slow	&	Yes	\\
EV Lac 	&	 22.08.2005 	&	53605.39147	&	330	&	300	&	187.0415	&	8.8183E+31	&	1.058	&	Fast	&	Yes	\\
EV Lac 	&	 22.08.2005 	&	53605.41460	&	195	&	105	&	61.7938	&	2.9134E+31	&	0.469	&	Slow	&	Yes	\\
EV Lac 	&	 22.08.2005 	&	53605.42241	&	75	&	30	&	20.6508	&	9.7361E+30	&	0.496	&	Slow	&	No	\\
EV Lac 	&	 22.08.2005 	&	53605.42467	&	105	&	75	&	24.9765	&	1.1776E+31	&	0.439	&	Slow	&	No	\\
EV Lac 	&	 22.08.2005 	&	53605.43113	&	45	&	30	&	12.4383	&	5.8642E+30	&	0.434	&	Slow	&	Yes	\\
EV Lac 	&	 23.08.2005 	&	53606.38421	&	45	&	30	&	8.4172	&	3.9684E+30	&	0.312	&	Slow	&	Yes	\\
EV Lac 	&	 23.08.2005 	&	53606.42194	&	45	&	30	&	7.6081	&	3.5869E+30	&	0.460	&	Slow	&	Yes	\\
EV Lac 	&	 23.08.2005 	&	53606.45404	&	120	&	105	&	30.8166	&	1.4529E+31	&	0.333	&	Fast	&	Yes	\\
EV Lac 	&	 23.08.2005 	&	53606.49226	&	45	&	15	&	1.6396	&	7.7299E+29	&	0.476	&	Slow	&	No	\\
EV Lac 	&	 23.08.2005 	&	53606.49851	&	30	&	15	&	3.5732	&	1.6846E+30	&	0.412	&	Slow	&	Yes	\\
EV Lac 	&	 23.07.2006 	&	53940.47444	&	60	&	30	&	7.9475	&	3.7470E+30	&	0.422	&	Slow	&	No	\\
EV Lac 	&	 23.07.2006 	&	53940.52270	&	480	&	375	&	113.1827	&	5.3362E+31	&	0.572	&	Fast	&	No	\\
EV Lac 	&	 23.07.2006 	&	53940.53726	&	120	&	90	&	16.4250	&	7.7438E+30	&	0.334	&	Slow	&	No	\\
EV Lac 	&	 29.07.2006 	&	53946.47599	&	1054	&	989	&	570.7213	&	2.6907E+32	&	1.991	&	Fast	&	No	\\
EV Lac 	&	 29.07.2006 	&	53946.54317	&	75	&	45	&	12.2062	&	5.7548E+30	&	0.394	&	Slow	&	No	\\
EV Lac 	&	 03.08.2006 	&	53951.53751	&	2898	&	2688	&	2579.1462	&	1.2160E+33	&	1.574	&	Fast	&	No	\\
EV Lac 	&	 04.08.2006 	&	53952.37326	&	125	&	65	&	12.4023	&	5.8473E+30	&	0.238	&	Slow	&	No	\\
EV Lac 	&	 04.08.2006 	&	53952.37413	&	30	&	20	&	3.6127	&	1.7033E+30	&	0.334	&	Slow	&	No	\\
EV Lac 	&	 04.08.2006 	&	53952.37911	&	20	&	10	&	2.8304	&	1.3344E+30	&	0.314	&	Slow	&	No	\\
EV Lac 	&	 04.08.2006 	&	53952.37934	&	20	&	10	&	2.6850	&	1.2659E+30	&	0.345	&	Slow	&	No	\\
EV Lac 	&	 04.08.2006 	&	53952.37980	&	30	&	20	&	4.2637	&	2.0102E+30	&	0.319	&	Slow	&	No	\\
EV Lac 	&	 07.08.2006 	&	53955.48708	&	90	&	60	&	14.2999	&	6.7419E+30	&	0.460	&	Slow	&	No	\\
EV Lac 	&	 07.08.2006 	&	53955.52215	&	90	&	60	&	7.6945	&	3.6277E+30	&	0.480	&	Slow	&	No	\\
EV Lac 	&	 07.08.2006 	&	53955.53095	&	120	&	90	&	28.0671	&	1.3233E+31	&	0.578	&	Slow	&	No	\\
EV Lac 	&	 08.08.2006 	&	53956.34904	&	60	&	45	&	13.7110	&	6.4642E+30	&	0.652	&	Slow	&	Yes	\\
EV Lac 	&	 08.08.2006 	&	53956.35441	&	120	&	105	&	38.2660	&	1.8041E+31	&	0.633	&	Fast	&	Yes	\\
EV Lac 	&	 08.08.2006 	&	53956.39257	&	1680	&	1590	&	900.4373	&	4.2452E+32	&	1.484	&	Fast	&	Yes	\\
EV Lac 	&	 08.08.2006 	&	53956.44677	&	30	&	15	&	8.5613	&	4.0364E+30	&	0.459	&	Slow	&	Yes	\\
EV Lac 	&	 08.08.2006 	&	53956.44989	&	150	&	75	&	23.5724	&	1.1114E+31	&	0.489	&	Slow	&	Yes	\\
EV Lac 	&	 08.08.2006 	&	53956.45215	&	75	&	45	&	16.8672	&	7.9523E+30	&	0.516	&	Slow	&	No	\\
EV Lac 	&	 08.08.2006 	&	53956.45285	&	30	&	15	&	8.7699	&	4.1347E+30	&	0.450	&	Slow	&	Yes	\\
EV Lac 	&	 08.08.2006 	&	53956.45510	&	195	&	105	&	56.8137	&	2.6786E+31	&	0.486	&	Slow	&	Yes	\\
EV Lac 	&	 12.08.2006 	&	53960.53098	&	2940	&	2100	&	1453.1544	&	6.8511E+32	&	0.821	&	Slow	&	No	\\
EV Lac 	&	 15.08.2006 	&	53963.40072	&	30	&	10	&	7.4410	&	3.5082E+30	&	0.578	&	Slow	&	Yes	\\
EV Lac 	&	 25.08.2006 	&	53973.49652	&	2330	&	2260	&	2288.6227	&	1.0790E+33	&	2.164	&	Fast	&	No	\\
EV Lac 	&	 05.09.2006 	&	53984.44139	&	360	&	315	&	253.0786	&	1.1932E+32	&	1.215	&	Fast	&	Yes	\\
EV Lac 	&	 05.09.2006 	&	53984.47750	&	135	&	60	&	35.1887	&	1.6590E+31	&	0.563	&	Slow	&	Yes	\\
EV Lac 	&	 05.09.2006 	&	53984.50997	&	30	&	15	&	8.6037	&	4.0564E+30	&	0.475	&	Slow	&	Yes	\\
EV Lac 	&	 08.09.2006 	&	53987.47180	&	90	&	45	&	19.8145	&	9.3418E+30	&	0.537	&	Slow	&	No	\\
EV Lac 	&	 08.09.2006 	&	53987.47389	&	75	&	30	&	24.6323	&	1.1613E+31	&	0.845	&	Slow	&	No	\\
EV Lac 	&	 15.09.2006 	&	53994.36716	&	70	&	20	&	19.0094	&	8.9622E+30	&	0.441	&	Slow	&	Yes	\\
EV Lac 	&	 15.09.2006 	&	53994.37087	&	320	&	150	&	59.0751	&	2.7852E+31	&	0.359	&	Slow	&	No	\\
EV Lac 	&	 17.09.2006 	&	53996.33563	&	435	&	390	&	82.6432	&	3.8963E+31	&	0.395	&	Fast	&	Yes	\\
EV Lac 	&	 17.09.2006 	&	53996.34258	&	360	&	150	&	65.9648	&	3.1100E+31	&	0.253	&	Slow	&	No	\\
EV Lac 	&	 17.09.2006 	&	53996.35595	&	195	&	135	&	35.6703	&	1.6817E+31	&	0.317	&	Slow	&	No	\\
V1054 Oph 	&	 11.06.2004 	&	53168.42510	&	140	&	80	&	13.5776	&	3.9221E+31	&	0.274	&	Slow	&	No	\\
V1054 Oph 	&	 14.06.2004 	&	53171.39288	&	650	&	600	&	440.3598	&	1.2720E+33	&	1.829	&	Fast	&	Yes	\\
V1054 Oph 	&	 14.06.2004 	&	53171.42147	&	180	&	130	&	10.7079	&	3.0931E+31	&	0.226	&	Slow	&	Yes	\\
V1054 Oph 	&	 14.06.2004 	&	53171.42563	&	3270	&	3190	&	1441.3629	&	4.1636E+33	&	1.239	&	Fast	&	Yes	\\
V1054 Oph 	&	 14.06.2004 	&	53171.46290	&	1780	&	1750	&	281.5233	&	8.1322E+32	&	0.359	&	Fast	&	Yes	\\
V1054 Oph 	&	 20.06.2004 	&	53177.40665	&	300	&	270	&	32.6526	&	9.4321E+31	&	0.355	&	Fast	&	Yes	\\
V1054 Oph 	&	 20.06.2004 	&	53177.42459	&	120	&	60	&	13.8765	&	4.0084E+31	&	0.230	&	Slow	&	No	\\
V1054 Oph 	&	 20.06.2004 	&	53177.43755	&	1000	&	520	&	59.0017	&	1.7043E+32	&	0.181	&	Slow	&	No	\\
V1054 Oph 	&	 04.07.2004 	&	53191.36033	&	340	&	90	&	43.0587	&	1.2438E+32	&	0.305	&	Slow	&	No	\\
V1054 Oph 	&	 04.07.2004 	&	53191.36334	&	1340	&	1280	&	329.4567	&	9.5168E+32	&	0.957	&	Fast	&	Yes	\\
V1054 Oph 	&	 04.07.2004 	&	53191.39505	&	1800	&	340	&	247.6494	&	7.1537E+32	&	0.194	&	Slow	&	No	\\
V1054 Oph 	&	 06.07.2004 	&	53193.34635	&	1070	&	960	&	429.5400	&	1.2408E+33	&	1.568	&	Fast	&	No	\\
V1054 Oph 	&	 06.07.2004 	&	53193.36302	&	1460	&	980	&	192.7904	&	5.5690E+32	&	0.180	&	Slow	&	No	\\
V1054 Oph 	&	 10.07.2004 	&	53197.39384	&	670	&	530	&	88.1072	&	2.5451E+32	&	0.502	&	Fast	&	No	\\
V1054 Oph 	&	 11.05.2005 	&	53502.41191	&	48	&	36	&	2.3143	&	6.6853E+30	&	0.151	&	Slow	&	No	\\
V1054 Oph 	&	 11.05.2005 	&	53502.41469	&	216	&	180	&	24.2944	&	7.0178E+31	&	0.365	&	Fast	&	Yes	\\
V1054 Oph 	&	 04.06.2005 	&	53526.42970	&	36	&	24	&	2.2093	&	6.3819E+30	&	0.122	&	Slow	&	No	\\
V1054 Oph 	&	 04.06.2005 	&	53526.44400	&	60	&	36	&	6.1596	&	1.7793E+31	&	0.175	&	Slow	&	Yes	\\
V1054 Oph 	&	 05.06.2005 	&	53527.40053	&	468	&	396	&	51.9875	&	1.5017E+32	&	0.271	&	Fast	&	Yes	\\
V1054 Oph 	&	 05.06.2005 	&	53527.40636	&	48	&	36	&	3.9567	&	1.1429E+31	&	0.192	&	Slow	&	No	\\
V1054 Oph 	&	 05.06.2005 	&	53527.45803	&	108	&	84	&	5.6538	&	1.6332E+31	&	0.152	&	Slow	&	Yes	\\
V1054 Oph 	&	 06.06.2005 	&	53528.38594	&	1596	&	708	&	160.4190	&	4.6339E+32	&	0.250	&	Slow	&	No	\\
V1054 Oph 	&	 06.06.2005 	&	53528.39942	&	1368	&	1008	&	168.3597	&	4.8633E+32	&	0.207	&	Slow	&	No	\\
V1054 Oph 	&	 06.06.2005 	&	53528.41775	&	288	&	276	&	31.2711	&	9.0331E+31	&	0.227	&	Fast	&	No	\\
V1054 Oph 	&	 13.06.2005 	&	53535.44630	&	2304	&	1260	&	280.2476	&	8.0953E+32	&	0.258	&	Slow	&	No	\\
V1054 Oph 	&	 13.06.2005 	&	53535.47463	&	1416	&	768	&	102.0127	&	2.9468E+32	&	0.182	&	Slow	&	No	\\
V1054 Oph 	&	 13.06.2005 	&	53535.48505	&	192	&	156	&	12.1108	&	3.4984E+31	&	0.216	&	Fast	&	Yes	\\
V1054 Oph 	&	 13.06.2005 	&	53535.49797	&	144	&	96	&	11.4981	&	3.3214E+31	&	0.314	&	Slow	&	Yes	\\
V1054 Oph 	&	 13.06.2005 	&	53535.50880	&	132	&	120	&	18.6571	&	5.3893E+31	&	0.395	&	Fast	&	No	\\
V1054 Oph 	&	 24.06.2005 	&	53546.42356	&	72	&	48	&	8.2024	&	2.3694E+31	&	0.374	&	Slow	&	Yes	\\
V1054 Oph 	&	 24.06.2005 	&	53546.48564	&	48	&	36	&	2.6777	&	7.7348E+30	&	0.227	&	Slow	&	No	\\
V1054 Oph 	&	 24.06.2005 	&	53546.48717	&	48	&	36	&	5.5471	&	1.6023E+31	&	0.295	&	Slow	&	No	\\
V1054 Oph 	&	 24.06.2005 	&	53546.48912	&	36	&	24	&	3.2846	&	9.4881E+30	&	0.242	&	Slow	&	No	\\
V1054 Oph 	&	 24.06.2005 	&	53546.49148	&	36	&	24	&	3.4479	&	9.9596E+30	&	0.266	&	Slow	&	No	\\
V1054 Oph 	&	 26.06.2005 	&	53548.40336	&	48	&	36	&	1.5149	&	4.3761E+30	&	0.151	&	Slow	&	No	\\
V1054 Oph 	&	 26.06.2005 	&	53548.47419	&	36	&	12	&	4.9893	&	1.4412E+31	&	0.421	&	Slow	&	Yes	\\
V1054 Oph 	&	 26.06.2005 	&	53548.48919	&	300	&	276	&	10.4920	&	3.0308E+31	&	0.313	&	Fast	&	Yes	\\
V1054 Oph 	&	 02.07.2005 	&	53554.36689	&	108	&	60	&	5.4493	&	1.5741E+31	&	0.212	&	Slow	&	Yes	\\
V1054 Oph 	&	 03.07.2005 	&	53555.33448	&	564	&	36	&	33.8135	&	9.7675E+31	&	0.209	&	Slow	&	No	\\
V1054 Oph 	&	 03.07.2005 	&	53555.33921	&	960	&	624	&	103.7402	&	2.9967E+32	&	0.231	&	Slow	&	No	\\
\enddata
\end{deluxetable}

\begin{table}
\begin{center}
\caption{For both fast and slow flares whose rise times are the same, the results obtained from both the regression calculations and the \textit{t-Test} analyses performed to the mean averages of the equivalent durations ($Log (P_{u})$) versus flare rise times ($Log (T_{r})$) in logarithmic scale are listed.\label{tbl-4}}
\begin{tabular}{lrr}
\\
\tableline\tableline
\textbf{Flare Groups :} & \textbf{Slow Flare} & \textbf{Fast Flare} \\
\tableline\tableline
\textbf{\textit{Data}} & & \\
Total Flare Number : & $30$ & $30$ \\
\tableline\tableline 
\textbf{\textit{Best Representation Values}} & & \\
Slope : & $1.109 \pm 0.127$ & $1.227 \pm 0.243$ \\
$y$ intercept when $x = 0.0$ : & $-0.581 \pm 0.226$ & $0.122 \pm 0.433$ \\
$x$ intercept when $y = 0.0$ : & $0.524$ & $-0.099$ \\
\tableline\tableline
\textbf{\textit{Mean Average of All Y Values}} & & \\
Mean Average : & $1.348$ & $2.255$ \\
Mean Average Error : & $0.092$ & $0.126$ \\
\tableline\tableline 
\textbf{\textit{Goodness of Fit}} & & \\
$r^{2}$ : & $0.732$ & $0.476$ \\
\tableline\tableline 
\textbf{\textit{Is slope significantly non-zero?}} & & \\
\textit{p-value} : & $<$  0.0001 & $<$  0.0001 \\
Deviation from zero? : & $Significant$ & $Significant$ \\
\tableline\tableline
\end{tabular}
\end{center}
\end{table}


\begin{thebibliography}{43}
\bibitem[Andrews (1982)]{And82} Andrews, A. D., 1982, IBVS, 2254, 1
\bibitem[Benz \& G\"{u}del (2010)]{Ben10} Benz, A. O. \& G\"{u}del, M., 2010, ARA\&A, 48, (in press)
\bibitem[Crespo-Chac et al. (2006)]{Cre06} Crespo-Chac\'{o}n, I., Montes, D., Garc\'{\i}a-Alvarez, D., Fern\'{a}ndez-Figueroa, M. J., L\'{o}pez-Santiago, J., Foing, B. H., 2006, \aap, 452, 987
\bibitem[Dawson \& Trapp (2004)]{Daw04} Dawson, B. \& Trapp, R. G., 2004, "In Basic and Clinical Biostatistics", The McGraw-Hill Companies Inc. Press, USA, p.61, p.134, p.245
\bibitem[Eggen (1965)]{Egg65} Eggen, O. J., 1965, Obs, 85, 191
\bibitem[Fleming et al. (1995)]{Fle95} Fleming, Th. A., Schmitt, J. H. M. M., Giampapa, M. S., 1995, \apj, 450, 401
\bibitem[Fossi et al. (1995)]{Fos95} Fossi, B. C. M., Landini, M., Fruscione, A., \& Dupuis, J., 1995, \apj, 449, 376
\bibitem[Gershberg (1972)]{Ger72} Gershberg, R. E., 1972, Astrophys. Space Sci. 19, 75
\bibitem[Gershberg et al. (1999)]{Ger99} Gershberg, R. E., Katsova, M. M., Lovkaya, M. N., Terebizh, A. V., Shakhovskaya, N. I., 1999, A\&AS, 139, 555
\bibitem[Gershberg (2005)]{Ger05} Gershberg, R. E., 2005, "Solar-Type Activity in Main-Sequence Stars", Springer Berlin Heidelberg, New York, p.53, p.191, p.192, p.194, p.211, p.325, p.360
\bibitem[Gordon \& Kron (1949)]{Gor49} Gordon, K. C., \& Kron, G. E., 1949, PASP, 61,210
\bibitem[Green et al. (1999)]{Gre99} Green, S. B., Salkind, N. J., Akey, T. M., 1999, "Using SPSS for Windows: Analyzing and Understanding Data", Upper Saddle River, N.J. ; London : Prentice Hall Press, P.50
\bibitem[Gurzadyan (1988)]{Gur88} Gurzadyan, G. A., 1988, \apj, 332, 183
\bibitem[Haisch et al. (1987)]{Hai87} Haisch, B. M., Butler, C. J., Doyle, J. G., \& Rodon\'{o}, M., 1987, A\&A, 181, 96
\bibitem[Hardie (1962)]{Har62} Hardie R.H., 1962, "in Astronomical Techniques", ed.W.A.Hiltner (Chicago: Univ. Chicago Press), 178
\bibitem[Haro \& Parsamian (1969)]{Har69} Haro, G., \& Parsamian, E., 1969, BOTT, 5, 45
\bibitem[Ishida et al. (1991)]{Ish91} Ishida, K., Ichimura, K., Shimizu, Y., Mahasenaputra, 1991, Ap\&SS, 182, 227
\bibitem[Joy (1947)]{Joy47} Joy, A. H., 1947, \apj, 105, 96
\bibitem[Joy \& Abt (1974)]{Joy74} Joy, A. H. \& Abt, H. A., 1974, \apjs, 28, 1
\bibitem[Kukarin (1969)]{Kuk69} Kukarin, B. V., 1969, "in General Catologue of Variable Stars", 3d ed., Moscow Sternberg Astronomical Institute
\bibitem[Landolt (1983)]{Lan83} Landolt, A. U., 1983, \aj, 88, 439
\bibitem[Landolt (1992)]{Lan92} Landolt, A. U., 1992, \aj, 104, 340
\bibitem[Leto et al. (1997)]{Let97} Leto, G., Pagano, I., Buemi, C. S., Rodon\'{o}, M., 1997, \aap, 327, 1114
\bibitem[Lippincott (1952)]{Lip52} Lippincott, S.L., 1952, \apj, 115, 582
\bibitem[Mavridis \& Avgoloupis (1986)]{Mav86} Mavridis, L. N. \& Avgoloupis, S., 1986, \aap, 154, 171
\bibitem[Mazeh et al. (2001)]{Maz01} Mazeh, T., Latham, D. W., Goldberg, E., Torres, G., Stefanik, R. P., Henry, T.J., Zucker, S., Gnat, O., Ofek, E. O., 2001, MNRAS, 325, 343
\bibitem[Moffett (1974)]{Mof74} Moffett, T. J., 1974, \apjs, 29, 1
\bibitem[Montes et al. (2001)]{Mon01} Montes, D., L\'{o}pez-Santiago, J., G\'{a}lvez, M. C., Fern\'{a}ndez-Figueroa, M. J., De Castro, E., Cornide, M., 2001, MNRAS, 328, 45
\bibitem[Motulsky (2007)]{Mot07} Motulsky, H., 2007, "In GraphPad Prism 5: Statistics Guide", GraphPad Software Inc. Press, San Diego CA, p.94, p.133
\bibitem[Osawa et al. (1968)]{Osa68} Osawa, K., Ichimura, K., Noguchi, T., \& Watanabe, E., 1968, Tokyo Astron. Bull., No 180
\bibitem[Oskanian (1969)]{Osk69} Oskanian, V. S., 1969, In: L. Detre (ed). "Non-Periodic Phenomena in Variable Stars", Proc. AUI Coll. No 4. Academic Press, Budapest. p.131
\bibitem[Pettersen et al. (1983)]{Pet83} Pettersen, B. R., Kern, G. A. \& Evans, D. S., 1983, \aap, 123, 184
\bibitem[Pettersen et al. (1984)]{Pet84} Pettersen, B. R., Coleman, L. A., Evans, D. S., 1984, \apj, 282, 214
\bibitem[Robrade et al. (2004)]{Rob04} Robrade, J., Ness, J. –U., and Schmitt, J. H. M. M., 2004, \aap, 413, 317
\bibitem[Rodon\'{o} (1978)]{Rod78} Rodon\'{o}, M., 1978, \aap, 66, 175
\bibitem[Rodon\'{o} (1986)]{Rod86} Rodon\'{o}, M., 1986, NASSP, 492, 409
\bibitem[Roques (1954)]{Roq54} Roques, P. E., 1954, PASP, 66, 256
\bibitem[Tokunaga (2000)]{Tok00} Tokunaga A. T., 2000, "Allen's Astrophysical Quantities", Fouth Edition, ed. A.N.Cox (Springer), p.143, p.479
\bibitem[Van de Kamp (1953)]{Van53} Van de Kamp, P., 1953, PASP, 65, 73
\bibitem[Van Maanen (1940)]{Van40} Van Maanen, A., 1940, \apj, 91, 503
\bibitem[Veeder (1974)]{Vee74} Veeder, G. J., 1974, \aj, 79, 702V 
\bibitem[Wall \& Jenkins (2003)]{Wal03} Wall, J. W. \& Jenkins, C. R., 2003, "In Practical Statistics For Astronomers", Cambridge University Press, p.79
\bibitem[Wilson (1954)]{Wil54} Wilson, R. H., Jr., 1954, \aj, 59, 132
\end{thebibliography}
\end{document}